\documentclass{egpubl}
\usepackage{eurovis2025}

\SpecialIssuePaper

\CGFccby

\usepackage[T1]{fontenc}
\usepackage{dfadobe}  

\usepackage{cite}  \BibtexOrBiblatex
\electronicVersion
\PrintedOrElectronic
\ifpdf \usepackage[pdftex]{graphicx} \pdfcompresslevel=9
\else \usepackage[dvips]{graphicx} \fi

\usepackage{egweblnk}
\usepackage{soul}
\usepackage{amsmath}

\graphicspath{{figs/}}

\usepackage{xcolor} 

\definecolor{linkColor}{HTML}{257E98}
\usepackage{soul}
\setuldepth{Berlin}
\newcommand\asLink[2]{\textcolor{linkColor}{\href{#1}{\ul{#2}}}}

\definecolor{bottleGreen}{RGB}{0, 106, 78} 

\newcommand{\etal}{et al.}
\newcommand{\etals}{et al.'s}

\newcommand{\ie}{{i.e.}}
\newcommand{\eg}{{e.g.,}}

\newcommand{\cf}{{cf.}}

\newcommand\todo[1]{}
\newcommand\am[1]{}
\newcommand{\sarah}[1]{}
\newcommand{\maggie}[1]{}
\newcommand{\al}[1]{}
\newcommand\eliza[1]{}

\newcommand{\secref}[1]{\hyperref[#1]{Sec.~\ref*{#1}}}
\newcommand{\appendixref}[1]{\hyperref[#1]{Appendix~\ref*{#1}}}
\newcommand{\figref}[1]{\hyperref[#1]{Fig.~\ref*{#1}}}
\newcommand{\eqnref}[1]{\hyperref[#1]{Eqn.~\ref*{#1}}}
\newcommand{\tabref}[1]{\hyperref[#1]{Table ~\ref*{#1}}}

\newcommand{\hlc}[2][yellow]{{\colorlet{foo}{#1}\sethlcolor{foo}\hl{#2}}}
\definecolor{surveyQuoteColor}{HTML}{008b1b}
\definecolor{quoteColor}{HTML}{C46299}
\definecolor{expertQuoteColor}{HTML}{008396}
\newcommand\qt[1]{\hlc[quoteColor!20]{``#1''}}
\newcommand\sqt[1]{\hlc[surveyQuoteColor!20]{``#1''}}
\newcommand\eqt[1]{\hlc[expertQuoteColor!20]{``#1''}}

\newcommand{\numSurvey}{83}
\newcommand{\numInterviewees}{11}

\newcommand{\linkToStudy}{\asLink{https://submitterpaper.github.io/study/Upset-Alttext-User-Survey/0}}

\newcommand{\condition}[1]{\textsc{#1}}
\newcommand{\condVis}{\condition{Vis}}
\newcommand{\condText}{\condition{Text}}
\newcommand{\condBoth}{\condition{Both}}

\newcommand{\sx}[1]{SP$_{#1}$}
\newcommand{\ix}[1]{P$_{#1}$}

\def\subsubsec#1
{\subsubsection{#1}}

\usepackage{enumitem}

\usepackage{listings}

\usepackage{booktabs}

\newcommand{\parahead}[1]
{\vspace{0.25em}\noindent \textbf{\textit{#1.}}}

\newcommand{\osf}{\asLink{https://osf.io/kbvs9/?view_only=d8ab7284d60347af923d79c253305720}{osf.io/kbvs9}}

\title{Accessible Text Descriptions for UpSet Plots}

\author[McNutt, McCracken, Eliza, \etal{}]
{\parbox{\textwidth}{\centering 
    A. McNutt$^{1}$\orcid{0000-0001-8255-4258},
    M.\,K. McCracken$^{1}$\orcid{0009-0006-5280-0546},
    I.\,J. Eliza$^{1}$\orcid{0000-0003-3087-5951}\thanks{The first three authors contributed equally} ,\\
    D. Hajas$^{2}$\orcid{0000-0002-2811-1197},
    J. Wagoner$^{1}$ \orcid{0009-0000-5053-2281},
    N. Lanza$^{1}$,
    J. Wilburn$^{1}$\orcid{0000-0002-7672-0798},
    S. Creem-Regehr$^{1}$ \orcid{0000-0001-7740-1118}, and
    A. Lex$^{1}$\orcid{0000-0001-6930-5468}
}
        \\
{\parbox{\textwidth}{\centering $^1$University of Utah, Salt Lake City, Utah\\
         $^2$Global Disability Innovation Hub, University College London
       }
}
}

\begin{document}

\teaser{
    \vspace{-2em}
    \includegraphics[width=0.9\linewidth]{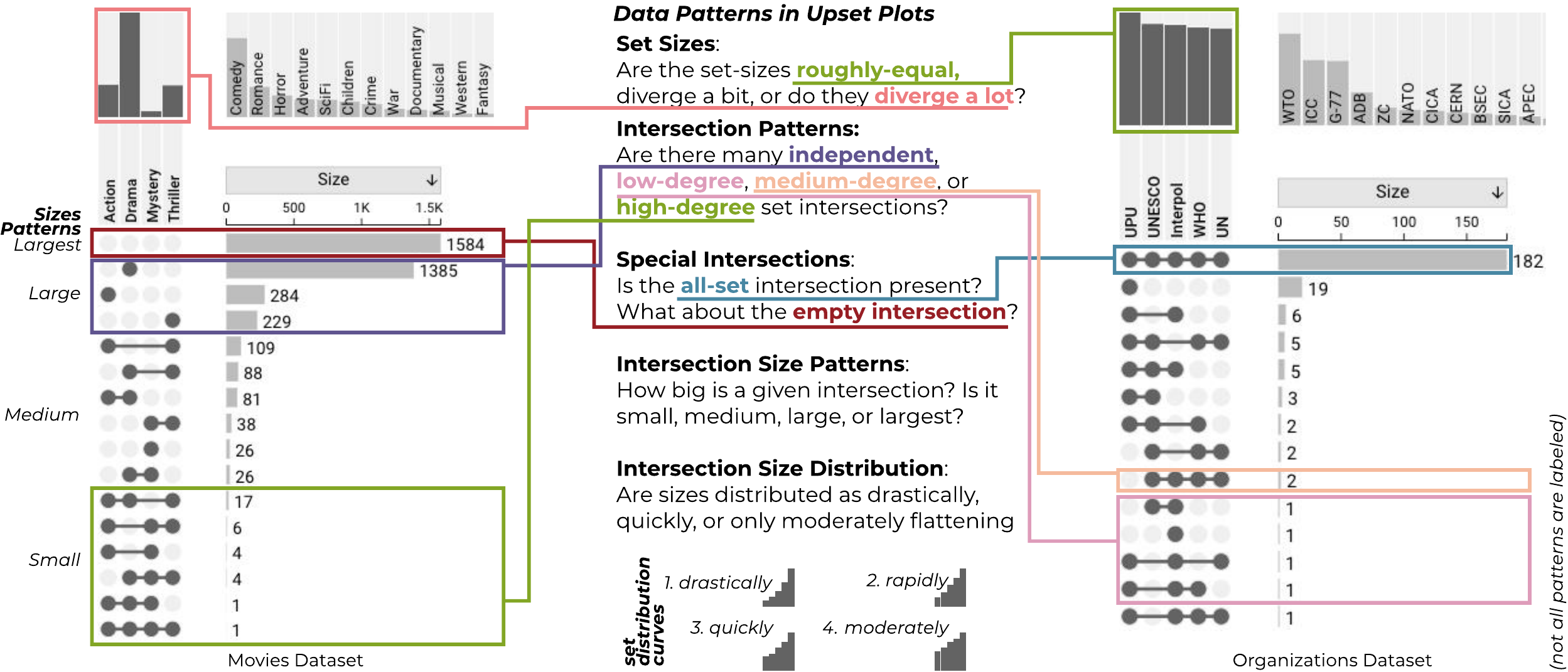}
    \centering
    \caption{
        UpSet plots are a prominent technique for visualizing set data. Despite their popularity, they remain inaccessible to blind or low vision (BLV) users.
        We seek to make this chart type accessible to all by developing a descriptive text-generation system. Here we show a subset of the descriptive data patterns that appears in their usage that enable our text descriptions.
    }
    \label{fig:patterns}
}

\maketitle
\begin{abstract}
    Data visualizations are typically not accessible to blind and low-vision (BLV) users.
    Automatically generating text descriptions offers an enticing mechanism for democratizing access to the information held in complex scientific charts, yet appropriate procedures for generating those texts remain elusive.
    Pursuing this issue, we study a single complex chart form: UpSet plots.
    UpSet Plots are a common way to analyze set data, an area largely unexplored by prior accessibility literature.
    By analyzing the patterns present in real-world examples, we develop a system for automatically captioning any UpSet plot.
    We evaluated the utility of our captions via semi-structured interviews with (N=\numInterviewees{}) BLV users and found that BLV users find them informative.
    In extensions, we find that sighted users can use our texts similarly to UpSet plots and that they are better than naive LLM usage.
\end{abstract}
\section{Introduction}

Being able to read and understand the content of information presented in data visualizations is crucial to fully grasp the content of scientific articles.
Yet, most visualizations are not accessible to blind or low-vision users (BLV)~\cite{butler_understanding_2017, shahira_assisting_2021}.
A common way to make images accessible is to provide text descriptions~\cite{jung_communicating_2022} that communicate the content of a chart to screen readers~\cite{zong_rich_2022}.
However, most scientific journals do not provide useful text descriptions for visualizations.
This lack of accessibility likely contributes to an under-representation of disabled individuals in STEM~\cite{nationalsciencefoundation_diversity_2023}.

The reasons for the lack of text descriptions in scientific publications are manifold---ranging from a lack of awareness, to technical challenges when providing text descriptions in scientific papers, to adding yet another burden to the onerous publishing process, and lack of knowledge on how to write useful text descriptions.
Automatic construction of text descriptions would seem then to offer a partial solution: reducing the burden of writing text descriptions and potentially forming a foundation for a human to refine.

Whereas pixel-based images are commonly analyzed and captioned using computer vision techniques (\eg{} on Facebook~\cite{wu_automatic_2017}), many chart forms are computationally represented and thus amenable to automated analysis and thus description.
Tang \etal{}~\cite{tang_vistext_2023} highlight that effective text descriptions should be specific to the chart and dataset type, yet, automatic approaches for data and chart aware charts are currently limited to simple charts, like one-dimensional line and bar charts. The construction of text for more bespoke or uncommon chart types has been identified as important next steps, although efforts so far have been limited~\cite{kim_explain_2023, li_altgeoviz_2024}.

Complicating these explorations is that usage and data patterns of complex charts are not as well as understood as those present in quotidian charts forms.
For instance, widely used plots, such as scatterplots, exhibit well-understood patterns, such as clusters or outliers, that algorithms can identify to extract information and potentially generate meaningful text descriptions.
Yet such patterns do not exist for exotic charts (such as necklace maps~\cite{speckmann_necklace_2010}), boutique charts (as in My Life with Long Covid~\cite{luipi_opinion_2023}), or even relatively well-known charts like tree maps~\cite{kim_explain_2023}.

We seek to improve the process of automatic descriptive text creation, by considering the specific case of UpSet plots~\cite{lex_upset_2014}.
UpSet plots offer an intriguing test case for text description-generation, as there has been substantial adoption (particularly in the biomedical domain with over 4000 citations to the two papers that introduce them~\cite{lex_upset_2014, conway_upsetr_2017}) making them a graphical form with potentially high impact if made more broadly accessible.
At the same time they consider a relatively uncommon data type; compared to the ubiquity of tabular data, set data is less frequently used---making them novel compared to prior designs.
They are hence an ideal exemplar of complex or scientific charts; offering fertile ground on which to explore alternative text design.

To explore text descriptions for UpSet plots, we developed a system that automatically generates descriptive texts.
To do so, we first identified common patterns in UpSet plots via a survey and qualitative analysis of published UpSet figures (\secref{sec:patterns}).
Based on the results of these patterns, we conducted a design process for effective text descriptions on UpSet, consistently collecting feedback from our blind coauthor, and drawing on insights from prior work~\cite{lundgard_accessible_2022}. We then implemented text generation that automatically detects the patterns we describe and generates the description.
We leverage a JSON-based description of the UpSet plot and the original dataset in the process. We implement our tool as a web-API, so that all kinds of implementations of UpSet plots can leverage our text generation, and demonstrate the process with two instances: an interactive, web-based version of UpSet and a Python library.
We design our texts generically, both to be generally useful and so that they can be manually adapted to specific situations.

To evaluate the quality of our results, we conducted a semi-structured interview study with (N=\numInterviewees{}) BLV participants (\secref{sec:interview}).
We find that the text descriptions gave participants a \qt{decent sense as to like what [UpSet Plots] might look like} and that they were informative and appropriate in length. We also received suggestions to improve the descriptions, for example, by adding a glossary and by structuring the text with bullet points.
More essentially, we find that participants were generally able to conduct at least simple analytic tasks using text descriptions that would be normally carried out using the visual representations.

We complement these findings with two contextualizing experiments, \secref{sec:experiments}.
First, we explored whether our descriptive texts contained the same level of data analytic functionality as visual UpSets. To do so, we conducted a (N=\numSurvey) crowd work study with sighted users. Although this does not show that our texts are equivalent to their visual representations, we do find that sighted users perform similarly well with text and visual versions in answering a series of factual questions. We also find that preference and correctness are highest for the combined text and visual condition, indicating that providing text descriptions for complex charts could have a ``curb-cut'' effect, benefiting both BLV and sighted users.
This also constitutes a minor contribution: this is the first user study of UpSet plots, highlighting that they can be understood with little training.
Then, we compared our texts to a naive usage of LLMs.
We find (via qualitative coding and expert reading from our blind coauthor) that they can produce usable texts, however the quality is variable and dependent on factors like presence of LLM training data relevant to the dataset.

By making it simpler to make scientific charts accessible we strive to bring about a world in which scientific communication can be understood and used by all.
Improving UpSet plot accessibility is a single step toward that larger goal.
Our materials (including our descriptive text generator) are available at \osf{}.
A live version of our implementation is available at \asLink{https://upset.multinet.app/}{upset.multinet.app}.

\section{Background and Related Work}

Our work builds upon prior research on accessibility in visualization and is considered in the context of UpSet plots.

\subsection{Accessibility in Visualization}

There is a growing interest in making visualizations accessible~\cite{kim_accessible_2021} using a wide range of approaches including sonification, physicalization, and alternative text.

Alternative or descriptive texts (sometimes called alt text) are a means to offer a nonvisual description of an image. These texts are typically accessed by an assistive technology called a screen reader (of which JAWS and VoiceOver are common examples) that vocalizes the descriptive text. This enables BLV users to understand what is being shown in purely visual media.

Despite the promise of these affordances, however, the lack of descriptive text for images and other web-based media is prominent.
For instance, L'Yi~\cite{lyi_digital_2024} find that data portals and journal websites typically do not have appropriate descriptive text.
Fortunately, there has been a steady sequence of work on improving this situation~\cite{wu_automatic_2017, gleason_twitter_2020, choi_visualizing_2019}.
For instance, Bigham \etal{}~\cite{bigham_webinsight_2006} develop a system (in 2006)  for automatically creating alt texts.
Morris \etal{}~\cite{morris_rich_2018} describe a design space of ways in which assistive technology could be improved to help users of screen readers.
In the process of generating meaningful alternative texts, understanding user preferences is important. Mack \etal{}~\cite{mack_designing_2021} show that users like summaries or content overviews in addition to more contextual descriptions.
Our work draws on these prior streams of descriptive text automation, but in the more specific context of visualization.

In visualization, text summaries can help users understand graphically presented data.
While there have been a variety of different approaches, these techniques usually center on descriptions of specific chart types (\eg{} line or bar charts)---which Jung \etal{}~\cite{jung_communicating_2022} highlight as an essential component of descriptive texts.
For instance, Moraes \etal{}~\cite{moraes_evaluating_2014} develop a system for automatically summarizing line charts.
Kim \etal{}~\cite{kim_explain_2023} develop a system for explaining a chart based on a schema, which allows description of new chart types like treemaps.
Most closely related to our work is AltGeoViz~\cite{li_altgeoviz_2024}, which develops descriptive text for a specific chart form (choropleth maps) outside of the common set of statistical graphics which alt text work usually focuses (as in Tang \etals{}~\cite{tang_vistext_2023} dataset).
We focus on a less well understood chart form (UpSet plots) and data type (sets), but strive toward the same goal of making all charts accessible.

Lundgard \etal{}~\cite{lundgard_accessible_2022} develop a four-rlevel semantic model of the types of content that appears in alternative texts.
Level 1 covers elemental visual components, such as the chart type or title.
Level 2 covers descriptive statistics, such as means, outliers, and so on.
Level 3 covers more complex trends identification or patterns.
Level 4 covers components related to the semantics of the data being represented, such as related to the domain or contextualizing events.
We make use of this model in the design of our descriptions.
Chintalapati \etal{}~\cite{chintalapati_dataset_2022} apply this model to analyze alt text usage in HCI publications---showing accessibility coverage to be lackluster.
Effective description appears to be contextual, and no one description will cover all cases---although automation efforts make exploration of information more personal via interaction.
For instance, Zong \etal{}~\cite{zong_rich_2022} develop a method for traversing high- and low-level trends represented and visualizations with screen readers (later reified as Olli~\cite{blanco_olli}).

Attempts have been made to use AI to automatically generate natural language summaries for charts \cite{obeid_chart-text_2020} and to create figure captions \cite{qian_formative_2020, qian_generating_2021}, but these summaries are limited to common chart types, a limitation we address in this paper.
Duarte \etal{}~\cite{duarte_autovizua11y_2024} use an LLM-based system to generate human quality alternative texts for simple statistical charts.
Interestingly, Mack \etal{}~\cite{mack_designing_2021} find that users starting from automated text craft lower quality descriptions than when starting from nothing, at least in the context of PowerPoint.
A component of our approach is that a human could adapt to their local context if they wanted, but we do not require them to do so as our text offers a functional starting point on its own.
Williams \etal{}~\cite{williams_supporting_2022} find that people often struggle to determine appropriate content for alternative texts---such as how to structure and compose their descriptions, particularly for figures with complex visual elements and relationships---highlighting the necessity for assistance in this area.

Outside of pure text descriptions there have been a number of different systems for chart accessibility.
Wang \etal{}~\cite{wang_enabling_2024} contribute a system that allows for nonvisual exploratory data analysis.
More broadly, Chundury \etal{}~\cite{chundury_understanding_2022} argue for additional sensory input modalities, such as touch (\ie{} physicalizations) and sound (\ie{} sonifications)---directions which have been broadly pursued.
Lundgard \etal{}~\cite{lundgard_sociotechnical_2019} explore physicalizations of classic visualizations, while Engel \etal{}~\cite{engel_user_2018} consider tactile versions of simple statistical graphics (like bar charts).
On the sonification side, Chart Reader~\cite{thompson_chart_2023} integrates sonification with alt text, whereas Erie~\cite{kim_erie_2024} builds a grammar of sonification (although it is not specific to accessibility).
Zong \etal{}~\cite{zong_umwelt_2024} continue this trend of blending accessibility-support modalities with their tool Umwelt, which combines visualization, sound, and text.
Elavsky \etals{} Data Navigator~\cite{elavsky_data_2024} broadens the scope by offering a generic accessibility interface for various modalities (including visualizations and physicalizations).
These tools are important for making charts more accessible; however, we deal with the more specific problem of improving text for scientific chart forms.

\subsection{UpSet Plots}

UpSets plots are a form of set visualization that highlights the sizes of various intersections with those sets, as well as related statistical distributions.
Originally developed by Lex \etal{}~\cite{lex_upset_2014}, UpSet plots are widely used across numerous implementations~\cite{lex_implementations_2024}.

\begin{figure}[t]
    \centering
    \includegraphics[width=\linewidth]{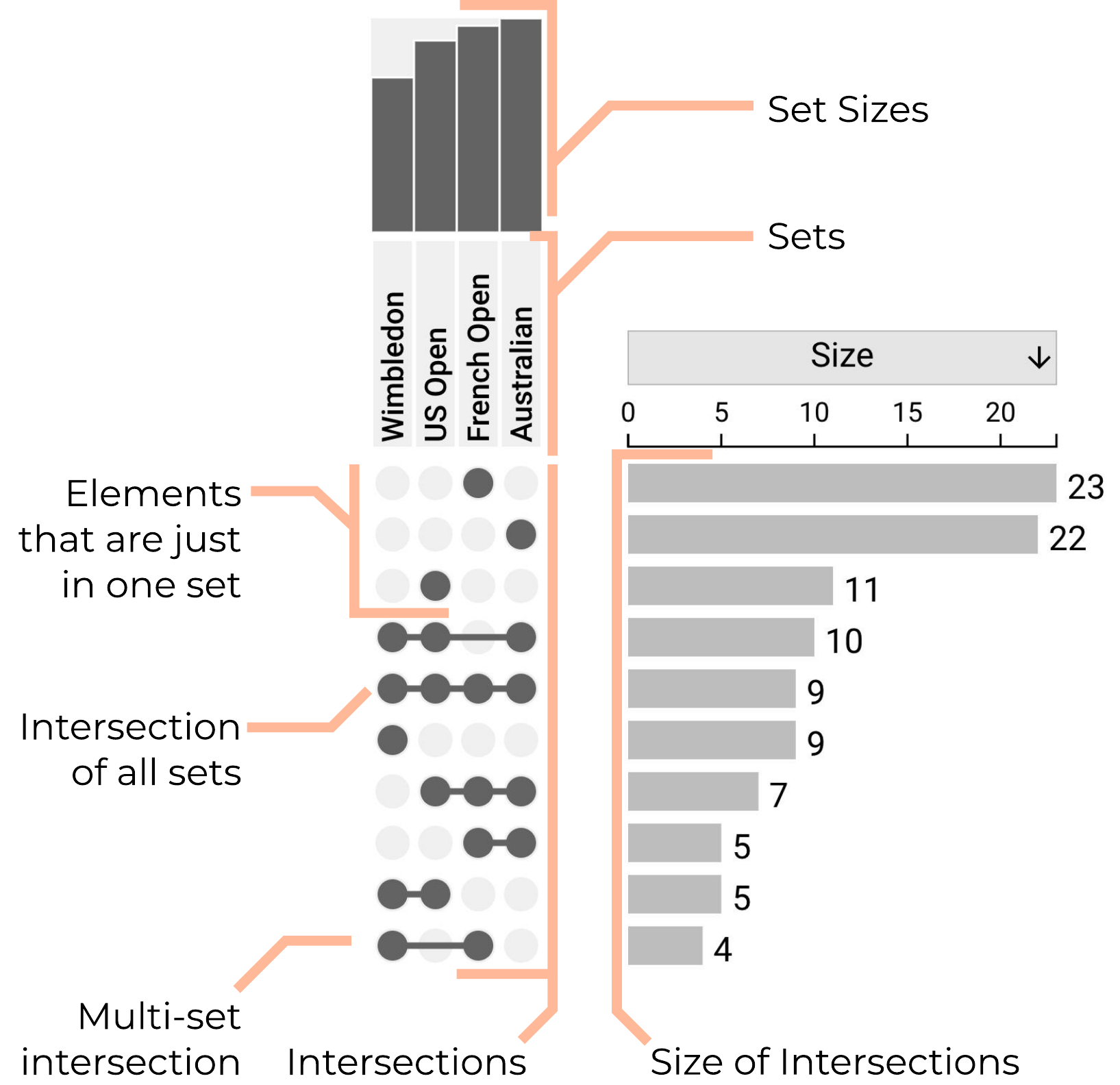}
\caption{
        An UpSet plot of the winners of major tennis tournaments, cropped to intersections with at least a size of 4. The sets are represented in the columns of the matrix; the set sizes as vertical bars; the size of the intersections as horizontal bars.
        Intersections are shown in the rows and identify the sets that intersect with filled in and connected circles.
        ``Intersections'' can be of only one set, here representing the players who have only won one tournament.
    }
    \label{fig:upset_explained}
    \vspace{-2em}
\end{figure}

\figref{fig:upset_explained} shows an example UpSet plot visualizing grand slam tennis winners. The \textit{elements} in the dataset are tennis players, the \textit{sets} are the four grand slam tournaments. The sets and intersections are shown in a matrix. Each column corresponds to a set, and bar charts on top show the size of the set. Each row corresponds to an \textit{intersection}: the filled-in cells show which set is part of an intersection. There are two special cases of ``intersections'': first, some elements are just in one set (tennis players that have only won one of the four Grand Slam tournaments). Second, an UpSet plot can also contain an ``empty intersection'' that contains elements that are not in any of the sets. In the tennis example, this would be players that never won one of the tournaments, which are not included in this dataset, and hence there is no empty intersection present.

The size of the intersections are displayed as bar charts to the right of the matrix. The matrix can be sorted in various ways to surface trends. For instance, a common way is to sort by size of the intersections.
Despite their prominence, there have been no previous user studies of UpSet plots. As an ancillary contribution, we provide experimental evidence that they can be used for simple tasks.

\section{Building Text Descriptions for UpSets}

We now describe our implementation of automated description generation for UpSet plots. To do so we first conducted a survey of extant UpSets to help identify the patterns necessary for effective description.
Based on the result of these patterns we design our descriptions through an iterative co-design process with a blind coauthor.
Finally, we synthesize these results as a system for generating automatic text descriptions.

\subsection{Surveying UpSet Patterns}
\label{sec:patterns}

UpSet plots, like any chart type, are more than just the specific marks that they use to encode information. Instead they have patterns wedded to those marks that allow verbal or textual description. For instance, in scatterplots there is a rich language of patterns, such as outliers, clusters, bimodal distributions, a variety of  scagnostics~\cite{wilkinson_scagnostics_2008} derived terms, and so on.
To form useful textual depictions of UpSet plots we need to be able to characterize the patterns present in them.
To do this, we surveyed published UpSet plots and conducted a qualitative analysis of them, yielding a collection of patterns, as summarized in \figref{fig:patterns}.

Seeking to understand real world usage, we gathered a collection of UpSets from published papers.
The most straightforward source of such plots can be found in the papers citing the UpSet papers---Lex \etals{}~\cite{lex_upset_2014} original description of the plot and Conway \etals{}~\cite{conway_upsetr_2017} UpSet R package.
As of late 2024, those papers have more than 4000 citations on Google Scholar, many of which use UpSet plots for data presentation.
For example, a paper on drug overdoses in the US uses UpSet plots to visualize overdoses involving multiple substances~\cite{friedman_xylazine_2022}.
As an analysis of all papers that cite these two papers and contain UpSet plots is infeasible due to the scale.
We instead drew a convenience sample that we iteratively grew until we reached saturation in our coding process, i.e., until we did not discover new types of patterns when adding additional papers to our corpus, a common method to determine when to stop adding data in qualitative research~\cite{corbin_basics_2014}.
To this end, we collected 80 UpSet plots from 41 published papers.
First, we searched for papers citing the original UpSet paper \cite{lex_upset_2014} in Google Scholar, resulting in 31 papers.
We then searched Google for the keyword ``UpSet Plot'' and found another 8 papers that we included in our sample. We collected plots from 2 more papers cited in our corpus.

Using this corpus, two authors developed a collection of patterns using an iterative qualitative coding process, combining apriori codes, based on the author's knowledge of UpSet plots and emergent codes discovered during the coding process~\cite{king_template_1998}.
First, one of them coded each plot individually based on the presence of the largest set in the largest intersection, the smallest sets in the smallest intersection (and vice versa), how the set sizes diverge, how the intersection sizes diverge, and the presence of all-set intersections (our apriori codes).
They then iteratively met to revise and alter the codes based on examples they saw (the emergent codes).
After two iterations, they reached a consensus and finalized the codes.

Based on this coding, we form five categories of \textbf{data patterns}, as well as a collection of \textbf{layout choices}.
The characteristics of a type of data pattern can often be described along a continuous variable. For example, correlation among two dimensions in a scatterplot can be described as an $R^2$ value. Since we intend to generate text descriptions based on these patterns, we discretize continuous variables to ordinal labels. For the $R^2$ example, a value of 0.9 could be mapped to ``highly correlated'', whereas a value of 0.4 could be considered ``moderately correlated''.
As a reminder, the data objects of interest in UpSets are sets and set intersections. We use UpSet plots for the Movies and Organization datasets (\figref{fig:patterns}) as examples. See \autoref{tab:heuristics} for the specific heuristics for each pattern.

\begin{description}[noitemsep, topsep=0pt]
    \item[Set Size Patterns.] Consider set size is essential when analyzing set intersections.
          If all sets in a dataset are of equal size, intersection sizes can be compared without considering set size. However, if set sizes diverge, an analysis of intersections also has to consider the size of the sets. For example, if 2/4 are very large while the other 2/4 are small, a large overlap between the large sets might visually dominate, even if those sets are only weakly correlated.
          The set sizes of the Movies dataset diverge a lot, as there are about 15 times as many Drama than Mystery movies; while in the Organizations dataset, the selected organizations are of roughly equal size.
          Most plots in our corpus were ``diverging a lot'', followed by ``moderately diverging'' and a few that were ``roughly equal''. Alarmingly, 18 of the plots we collected did not show set size, which has the potential to mislead viewers.

    \item[Intersection Patterns.]
          There are $2^n$ possible intersections in a set dataset, where $n$ is the number of sets. For most set datasets, many of the possible intersections are empty. Moreover, different data types exhibit vastly different patterns with respect to which intersections exist. The Movies dataset, for example, contains many movies with just one or two genres, where intersections between only a few sets dominate. The Organizations dataset, on the other hand, has many countries that are part of the same organizations, resulting in several nonempty intersections that involve many organizations.

    \item[Special Intersection.]
          We treat two intersection patterns as special: the ``all-set intersection'', and the ``empty intersection'' (containing the elements that are in no set but are still in the dataset). We treat these two patterns distinctly, because we always report on their presence (and their size), while we rely on size attributes of other intersection patterns to determine whether to explicitly call them out. Both the Movies and the Organizations datasets contain the all-set intersection, while only the Movies dataset contains the empty intersection.

    \item[Intersection Size Patterns.]
          Different types of intersection patterns can be associated with different intersection sizes. We report on the association of intersection patterns with respect to their size.
          We distinguish between small, medium, large, and largest sizes.
          For example, in the Movies dataset, the ``empty intersection'' is largest, the ``independent sets'' are large, and the ``high-degree'' intersections are small.

    \item[Intersection Size Distribution]
          Finally, we classify the distribution of the intersection sizes based on which of several functions they best fit (when sorted by magnitude).
          These include drastically flattening and rapidly flattening (exponential curves of different degrees), quickly flattening (quadratic curves), or steadily flattening (linear curves).
          For instance, Movies is rapidly flattening, while Organizations flattens drastically.
          A constant curve is possible, but we saw no evidence of such a dataset.
\end{description}

We also coded for \textbf{layout and presentation}.
Most notably, we found that UpSet plots are commonly \textit{sorted} by the size of the intersection, although other sorting strategies (\eg{} by degree or attribute) are also possible.
In our corpus, most plots were sorted by intersection size, followed by degree.
We also found a few instances of attribute-driven sortings (\eg{} by a group first, and by intersection size second)~\cite{friedman_xylazine_2022}, or by a measure of deviation from expected intersection size~\cite{lu_comparative_2022}.
We frequently found custom \textit{highlights and annotations}, and supplementary plots that show attributes or Venn diagrams to explain the intersections.
They were \textit{oriented} either horizontally or vertically, but most were horizontal.

The original UpSet implementation~\cite{lex_upset_2014} also introduces set filters, aggregations, and queries. We found only one instance of \textit{aggregation} in our corpus. The original UpSet enables users to interactively select which sets to include in the dataset, but we found no occurrence of sets not included in the plot in any of the figures, which makes sense given that we drew our figures from published papers, and choosing sets is only useful in an exploratory stage.

\begin{table}[t]
    \centering
    \caption{The pattern types and the specific names in categorization of UpSet usages. In the supplement we analyze application of these patterns to a variety of datasets. $\cap$ = intersection.
    }
    \small
    \begin{tabular}{p{0.17\linewidth}p{0.26\linewidth}p{0.42\linewidth}}
        \textbf{Type}                     & \textbf{Name}                  & \textbf{Heuristic}                                          \\
        \toprule
        Set Size
                                          & Diverging a lot                & Vary by $>30\%$                                             \\
                                          & Moderately Div.                & Vary by $10\% - 30\%$                                       \\
                                          & Roughly Equal                  & Vary by  $<10\%$                                            \\

        \midrule

        $\cap$ Patterns                   & Independent sets               & $\cap$s containing only 1 set                               \\
                                          & Low-degree $\cap$s             & $\cap$s containing $2-3$ sets                               \\
                                          & Medium-degree $\cap$s          & $\cap$s containing $3-\frac{n}{2}$ sets                     \\
                                          & High-degree $\cap$s            & $\cap$s containing $\frac{n}{2}$ to $n$ sets                \\
        \midrule

        Special $\cap$s                   & All Set $\cap$                 & Check for presence                                          \\
                                          & Empty Set                      & Check for presence                                          \\
        \midrule

        $\cap$ Size\newline Patterns      & Small                          & $\cap$ size smaller than median of intersection sizes       \\
                                          & Medium                         & Size between median and median + 1.5* IQR                   \\
& Large                          & Bigger than median + 1.5* IQR but not max                   \\
                                          & Largest                        & Biggest intersection size                                   \\
        \midrule

        $\cap$ Size \newline Distribution & Drastically\newline Flattening & $\cap$ sizes follow an exp. curve with  high $\beta$ (>0.8) \\
                                          & Rapidly\newline Flattening     & Sizes follow an exp. curve  with low $\beta$ (<0.8)         \\
                                          & Quickly Flattening             & Sizes follow a quadratic curve
        \\
                                          & Steadily Flattening            & Sizes follow a linear curve                                 \\
        \bottomrule
    \end{tabular}
    \label{tab:heuristics}
    \vspace{-1em}
\end{table}

\subsection{Text Description Design}

Next, we used these patterns, along with general properties of the UpSet plot, to design text descriptions for UpSet plots.
We approach this as a design problem, as the structure of text is just as much a UI as a traditional GUI~\cite{yalanska_ux_2019}.
At every step of the process, we elicited feedback from our blind coauthor to ensure that our text-design is useful for the target audience. We used Lundgard and Satyanarayan's~\cite{lundgard_accessible_2022} semantic levels of context to design our text descriptions because they constitute a validated framework.
We designed our text so that it could act as a launching pad for humans to improve instead of a final text to use.

\parahead{Design Process}
Our design process involved an iterative cycle of prototyping, abstracting, and refining.
We started by having one author manually write three text descriptions for representative UpSet plots.
They followed a template based on Lundgard and Satyanarayan's~\cite{lundgard_accessible_2022} semantic levels of context.
This diversity of plots allowed us to explore how different patterns would be reported.
We then started iterating on these descriptions also with our blind coauthor.
At the same time, we looked for commonalities in descriptions with an eye toward generalizing.
After reaching an initial stable point, we implemented a basic version of the text generation.
However, this revealed unanticipated issues with the text.
For example, our blind coauthor noted that a dataset with complex set labels (\eg{} \verb+ALL_X_citri_subsp_citri_UI6_NZ_CP008992+) resulted in hard-to-follow utterances by a screen reader.
We decided to (a) encourage users of the API to use human-readable set names, and (b) for cases when we had to deal with such strings to extract a short human-readable string---`citri sub' in this case.
In this phase, we also investigated three text lengths. We eventually abandoned one of the sizes, preferring two based on feedback from our collaborator, as it was difficult to distinguish between the medium and highly verbose levels when using a screen reader.

\begin{figure}[t]
    \centering
    \includegraphics[width=0.85\linewidth]{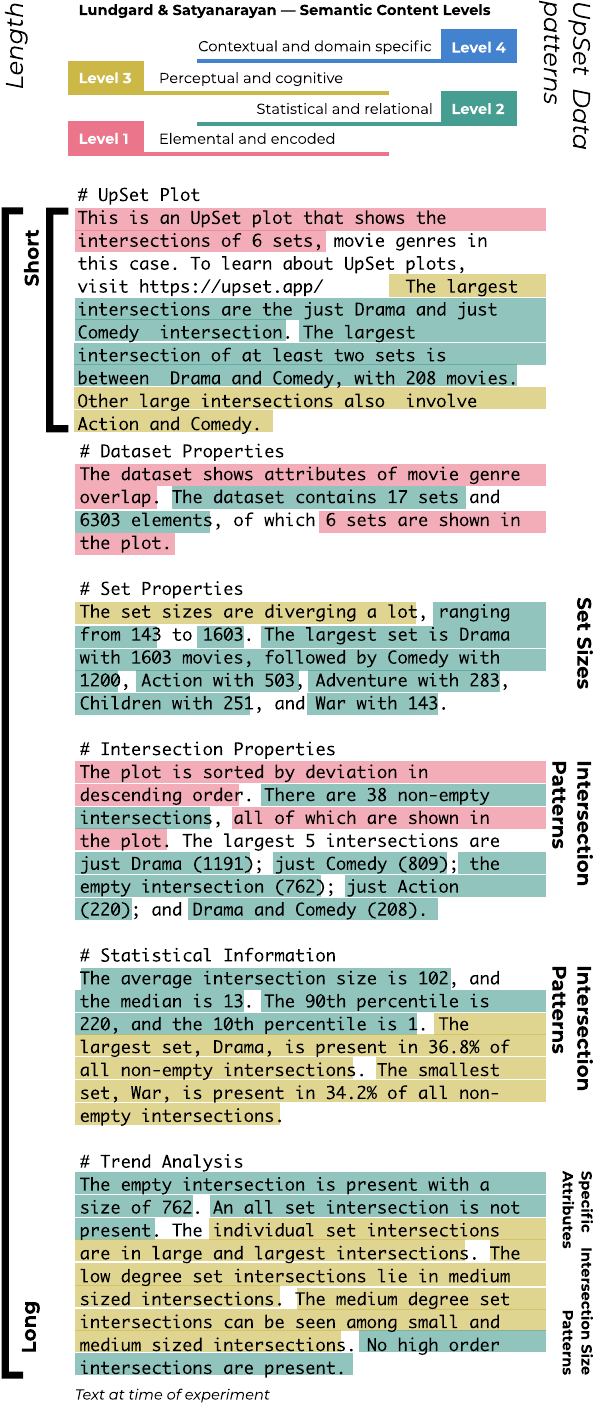}
\vspace{-3mm}
    \caption{
        Our short and long text descriptions annotated with Lundgard and Satyanarayan's~\cite{lundgard_accessible_2022} semantic text model.
    }
    \label{fig:set-example}
    \vspace{-10mm}
\end{figure}

\parahead{Designed Text}
\figref{fig:set-example} shows the resulting long description next to a web-based UpSet implementation.
The \textit{Short Description} is a brief snippet that contains the most salient trend and that acts as a high-level overview of plot, akin to image alt text.
The \textit{Long Description} is an extended description (including the short description as the introduction) designed to communicate similar amounts of information as the chart. When available, we leverage human-provided labels of the sets and items to generate more natural text. For example, in the sentence ``\texttt{The largest intersection is between Drama and Comedy, with 208 items.}'' we replace ``\texttt{items}'' with ``\texttt{movies}''.

The long description includes Introduction, Dataset Properties, Set Properties, Intersection Properties, Statistical Information, and Trend Analysis. The first four sections have basic information about the type of the plot, how many sets are involved, how many intersections are seen, the count of visible sets, the title of the sets, sorting order, largest intersection name, the largest five to ten intersections, etc. For the statistical information section, we compute a variety of metrics including percentile information, the percentage of the presence of the largest set and smallest set in all the intersections, the average value, and the median divergence of the intersections.
Finally, the trend analysis section describes trends that we determine to be present based on our observed data patterns. Examples of both long and short descriptions for several datasets are available at \asLink{https://upset.multinet.app/}{upset.multinet.app}.

In designing this description we expected that we are unable to adequately generate Level 4 content (contextual and domain specific) because it requires knowledge of the data and context in which the chart is being used that is outside of the purview of our system.
For instance, observations such as ``The explosion of \$200M+ superhero film budgets in the mid-2010s demonstrates the impact of Marvel's shared universe model, fundamentally altering how studios approach franchise development'' would be difficult to generate automatically.
We encourage authors of UpSet plots to edit the text descriptions and provide such context.
Subsequent work could explore automation of this semantic description step through LLMs, however, as we discuss in \secref{sec:experiments}, this can lead to hallucination and subtle misinterpretation.

\parahead{Generating Text}
Our text descriptions are generated using simple text templates, one version for the short form description and another for the long.
The structure of these templates is alluded to in \figref{fig:set-example}. For instance, in the long version there are sections for \texttt{Set Properties}, \texttt{Intersection Properties}, and so on.
The content of these sections is formed from a descriptive JSON-based standard that summarizes the configuration of an UpSet plot---such as how Vega-Lite programs summarize the structure of the resulting visualization.
It consists of the data, visible sets, plot direction, and sorting method---as well as aspects like aggregation and attribute configuration that we do not currently leverage for our descriptions.
In addition to structural metadata, statistical information (such as ``\texttt{The dataset contains 17 sets and 6303 elements}'') is generated by analyzing the input dataset following our observed patterns using standard analytic techniques.
In essence, our descriptions are simple functions that take in a configuration and dataset and return markdown containing the description.
We use this design, rather than one more closely woven into a particular tool, so that any tool that generates UpSet plots might be instrumented to create these text descriptions.

\begin{table*}[t]
    \centering
    \small
    \caption{Participant demographics. Severe low vision is classified as a Snellen visual acuity between 20/200 and 20/500, profound low vision is classified as a Snellen visual acuity between 20/500 and 20/1000 \cite{americanoptometricassociation_low_2024}. IP = degree in progress; $\bar{X}$ is mean; $Y$ is mode.}
    \begin{tabular}{cccccccc}

        \textbf{pID} & \textbf{Age} & \textbf{Gender} & \textbf{Daily Hours} & \textbf{Screen-Reader} & \textbf{Described Level} & \textbf{Vision-Loss}  & \textbf{Highest Level} \\
                     &              &                 & \textbf{on Computer} & \textbf{(Years Exp.)}  & \textbf{of Vision}       & \textbf{Level}        & \textbf{of Education}  \\
        \toprule
        \ix{1}       & 49           & Male            & 8                    & VoiceOver (39)         & Severe low vision        & Lost vision gradually & Bachelor's             \\
        \ix{2}       & 71           & Female          & 12                   & VoiceOver (1)          & Severe low vision        & Lost vision suddenly  & High School            \\
        \ix{3}       & 24           & Female          & 4                    & JAWS (16)              & Light perception         & Lost vision suddenly  & Associate's (IP)       \\
        \ix{4}       & 35           & Male            & 8                    & JAWS (16)              & Light perception         & Lost vision suddenly  & High School            \\
        \ix{5}       & 24           & Female          & 4                    & VoiceOver (10)         & Light perception         & Blind since birth     & High School            \\
        \ix{6}       & 27           & Female          & 6                    & JAWS (14)              & No residual vision       & Lost vision suddenly  & Master's               \\
        \ix{7}       & 34           & Female          & 10                   & JAWS (12)              & Light Perception         & Lost vision gradually & Master's               \\
        \ix{8}       & 32           & Female          & 4                    & NVDA (28)              & No residual vision       & Blind since birth     & Associate's            \\
        \ix{9}       & 27           & Female          & 14                   & VoiceOver (8)          & No residual vision       & Blind since birth     & Bachelor's             \\
        \ix{10}      & 23           & Male            & 6                    & VoiceOver (12)         & Severe low vision        & Lost vision gradually & Bachelor's (IP)        \\
        \ix{11}      & 21           & Male            & 7                    & VoiceOver (8)          & Profound low vision      & Lost vision gradually & Bachelor's (IP)        \\
        \midrule
                     & $\bar{X}=$   & 7/11 F          & $\bar{X}=7.5 \pm 3$  & $Y$=VoiceOver,         & $Y$=                     & $Y$=suddenly(4),      & $Y$=                   \\
                     & $33 \pm 15$  &                 &                      & ($\bar{X}=15 \pm 10$)  & Light perception (4)     & gradually (4)         & Bachelor's (4)         \\
        \bottomrule
    \end{tabular}
    \label{tab:demographics}
    \vspace{-1em}
\end{table*}

\parahead{Implementation}
To support these goals we package our system as an open-source library \asLink{https://github.com/visdesignlab/upset-alt-txt-gen}{github.com/visdesignlab/upset-alt-txt-gen}.
To test the effectiveness and abstraction of our library, we explored using it to describe charts generated by our own interactive UpSet tool (described below) and UpSetPlot~\cite{nothman_upsetplot_2019}.
We had to only minimally modify UpSetPlot to generate our JSON standard, so that we could use our library to generate text descriptions.
To support additional libraries, we provide a REST API that accepts our standard and a dataset, and returns a text description. In this way, other versions of UpSet (including our web-based implementation) can use our library to generate text descriptions independent of the programming language used.

We also built a reference tool for creating UpSet plots (\asLink{https://upset.multinet.app/}{upset.multinet.app}) that includes a user-provided title and caption, and our text descriptions (see supplemental video).
Our tool enables users to edit the text description so that it can be tailored to the specific plot and dataset, while still providing access to the generated text.
The description dynamically updates as the plot is manipulated, e.g., adding or removing sets or changing the sorting.
To provide further access to BLV users, we include tabular versions of the dataset.
It is designed to be screen-reader accessible and was validated with our blind coauthor.

\section{Evaluation: Do BLV Users Find Our Descriptions Useful?}
\label{sec:interview}

With our ability to automatically generate text descriptions of UpSet plots now in hand, we want to ensure that these descriptions are useful to our target audience---namely BLV users.
In particular, we sought to identify whether our text descriptions sufficiently portray the quantitative information of an UpSet plot to users who cannot see the corresponding visualization. To address this, we conducted a semi-structured interview study with BLV participants.

We find that, generally, people were able to use our descriptions to answer analytic questions and liked their overall structure, but they pointed out potential improvements (\eg{} adding bullet points to support navigation).

\subsection{Procedure}

Interviews took place over Zoom using a pair-interview~\cite{akbaba_two_2023} technique in which two research team members were present for all interviews.
One interviewer acted as the driver who was primarily responsible for interacting with the participant and asking the prescribed interview questions. The second interviewer actively listened to the discussion and offered follow-up questions.
Interviews lasted in total an average of $50\pm 14$ minutes.
After obtaining consent, we asked demographic questions for $8\pm 3$ minutes (the results of which are shown in \autoref{tab:demographics}). We conducted a pilot with our blind coauthor as a process check.

\parahead{Participants}
We recruited screen-reader users by circulating an IRB-approved flyer distributed at the University of Utah Moran Eye Center and social media groups. Subjects were eligible to participate if they were 18 years of age or older, proficient in English, legally blind (\eg{} a visual acuity of 20/200 or greater), and use screen readers daily (as we consider users who regularly use screen-readers as the target audience of our text descriptions).
Of the 11 participants, 9 completed the interview on personal laptops and two completed the interview on a smartphone.
Participants received a \$50 Amazon gift card as compensation.
All participants provided informed consent for the procedures approved by our institution's IRB and to have the audio of the interview recorded and automatically transcribed. Transcriptions were manually corrected post hoc.
We refer to our interview Participants as \ix{X} and \qt{quote them}.

\parahead{Training}
Next, we directed participants to a training page sent via Zoom chat or email (\asLink{https://vdl.sci.utah.edu/upset-alttext-example/}{vdl.sci.utah.edu/upset-alttext-example}).
This page includes a basic description and guide to interpretation for UpSet Plots.
We then asked participants to share their screen and audio (screen reader output) to observe which part of the text they were interacting with---which we continued for the rest of the interview.
Participants were allowed to reread the training and ask clarifying questions.
Training lasted an average of $9\pm 2$ minutes.

\parahead{Text Description}
Next, participants opened a new web page with an example UpSet text description.
We used a Covid dataset showing co-occurrences of major symptoms (see the appendix for details).
We allowed participants to read the text as many times as necessary to understand the data.
We then asked participants what they learned from the description.
Specific questions included:
Can you describe what you learned about the dataset in your own words?
How did this dataset increase your understanding of COVID-19 symptoms?
We then asked for participants' feedback, including: What was difficult for you to understand about the dataset? What would have been helpful to provide additionally? Do you have comments on the style of the text description? We also asked factual questions, such as: How many sets are shown? What is the largest intersection? The part of the interview lasted $21\pm 7$ minutes.

\parahead{UpSet Comparison}
For participants who had some residual vision, we also
provided the corresponding COVID-19 symptoms UpSet visualization at this point.
We asked these participants if their mental model from the alt-text matched the chart and solicited additional feedback after they had seen the corresponding visualization.

\subsection{Results}
We now present the findings summarized by theme.

\parahead{Experience with Screen Readers}
To make text accessible to BLV users, it is important to understand the context in which screen readers are typically used. Most participants frequently use more than one screen reader, depending on the electronic device they use. For the interview,  the most commonly used screen reader was Apple's VoiceOver, followed by JAWS (see \autoref{tab:demographics}).
Prior experience with text descriptions for visualizations varied drastically.
\ix{1}, \ix{4}, \ix{7}, \ix{10} mentioned encountering  visualization text descriptions on a daily basis, \ix{2}, \ix{5}, \ix{9}, \ix{11} did so at least once a week, while \ix{3}, \ix{6}, \ix{8} encountered them less than once a week.
Most participants encountered visualizations in news sources,
for instance, \ix{9}, noted \qt{I come across them in news articles and magazines sometimes}.
Text description for  visualizations were common at school and work settings.
\ix{7} explained \qt{I make use of graphs and charts a lot, work related}. Similarly, \ix{3} recalled \qt{I am taking a human biology course and that involves some graphics and data charts.}

\parahead{Interpretation of the Text Description}
Participants sometimes misinterpreted and incorrectly used unfamiliar factual terms.
For instance, \ix{2} highlighted the difficulty of some technical terms: \qt{I think the word intersection might have been in this case an intersection is a person and a symptom together.}
Correctly understanding the term intersection was also demonstrated by using the term correctly without prompting. For instance, \ix{3} mentioned, \qt{I have had COVID a couple of times myself. I could put myself in the intersection [of symptoms] I've experienced.} Echoing Wang \etals{}~\cite{wang_how_2024-2} finding that BLV users draw on personal experience to understand charts.
Another mistake was misidentifying the largest intersection. To wit, some participants (4/11) were unable to correctly identify anosmia and fatigue as the largest intersection.

Participants reported a mixed sense of understanding the COVID-19 data.
Those with a strong grasp on it compared their knowledge with how a sighted person might perceive the UpSet plot.
For example, \ix{3} noted that \qt{I feel like I have a pretty decent sense as to like what this might look like if someone were to look at [the plot] visually.}
Others comments included having a good mental image of the corresponding UpSet plot (\ix{10}) and a better understanding of the co-occurrence of COVID-19 symptoms (\ix{11}).
Those with a more limited understanding still felt they got the important information.
\ix{8} noted that \qt{if I needed specific information, I could get it from [the text description]} but had a difficult time visualizing the associated chart, a position echoed by \ix{6} and \ix{7}.

\parahead{Feedback on the Text Descriptions}
Participants provided valuable feedback on improving our text descriptions, emphasizing clarity and accessibility. A common suggestion was to include the glossary of terms from the Introduction page within the description, such as at the top or bottom.
For instance, \ix{4} complained that \qt{I don't know the meanings of some of the words like elements.}
Providing a glossary simultaneously (alongside the description) may also reduce misinterpretations of terms.
\ix{10} noted concerns with terminology depending on the text description writer's training as \qt{someone could write the text description and put it a different way}---highlighting the role of expertise and domain knowledge for both the writer (and, we add, the reader).

Participants, overall, felt the length of the description was appropriate. Many noted that it was concise (\ix{2}, \ix{4}) and descriptive (\ix{3}, \ix{9}).
\ix{5} explained \qt{even though there is lots of very complex data, the numbers really help me understand things better.}
Several participants commented on the amount of numbers in the description. For example, \ix{11} recalled being unsure about the meaning of the numbers but stated, \qt{it was clear after the second read.}
Participants also noted that numbers are fully read out by screen readers. \ix{2} explained that the number \qt{281} might be skimmed over when reading visually while a screen reader fully articulates it as \qt{two hundred and eighty-one}.
\ix{3} similarly noted that fully articulated numbers \qt{takes me a little bit longer to go back and read them.} We also observed that several participants slowed down the speech rate of their screen readers to receive the information at a slower pace.
This suggests that numbers could be summarized with fewer significant figures with by a high fidelity table of numbers elsewhere (akin to the tables often accompanying visualizations~\cite{zong_rich_2022}).

Consistent with the feedback from our blind coauthor, all participants valued having headings in the alt-text.
Participants stressed that headers provide an array of benefits to screen readers.
For instance, organizing information by headings is \qt{easier to navigate with screen readers} (\ix{5}) and \qt{helps you get right where you need very easily} (\ix{8}).
\ix{7} noted that the headings \qt{helped with how I was processing what I was reading.}
Several participants suggested adding bullet points to the alt-text as an additional organizational tool.
Others indicated that adding bullet points alongside the existing headings would help with navigating long texts (\ix{8}, \ix{11}), breaking up lengthy sentences (\ix{2}), and organizing technical or number-intensive sections, such as statistical information (\ix{6}, \ix{10}).
In addition to benefiting screen readers, organizing the information this way may be useful for other accessibility tools.

Based on these results, we updated  text description to  the results bullet-point and added a glossary, as in Supplementary Figure~\ref{fig:bullets-glossary}.

\begin{figure*}[ht]
    \centering
    \includegraphics[width=0.9\linewidth]{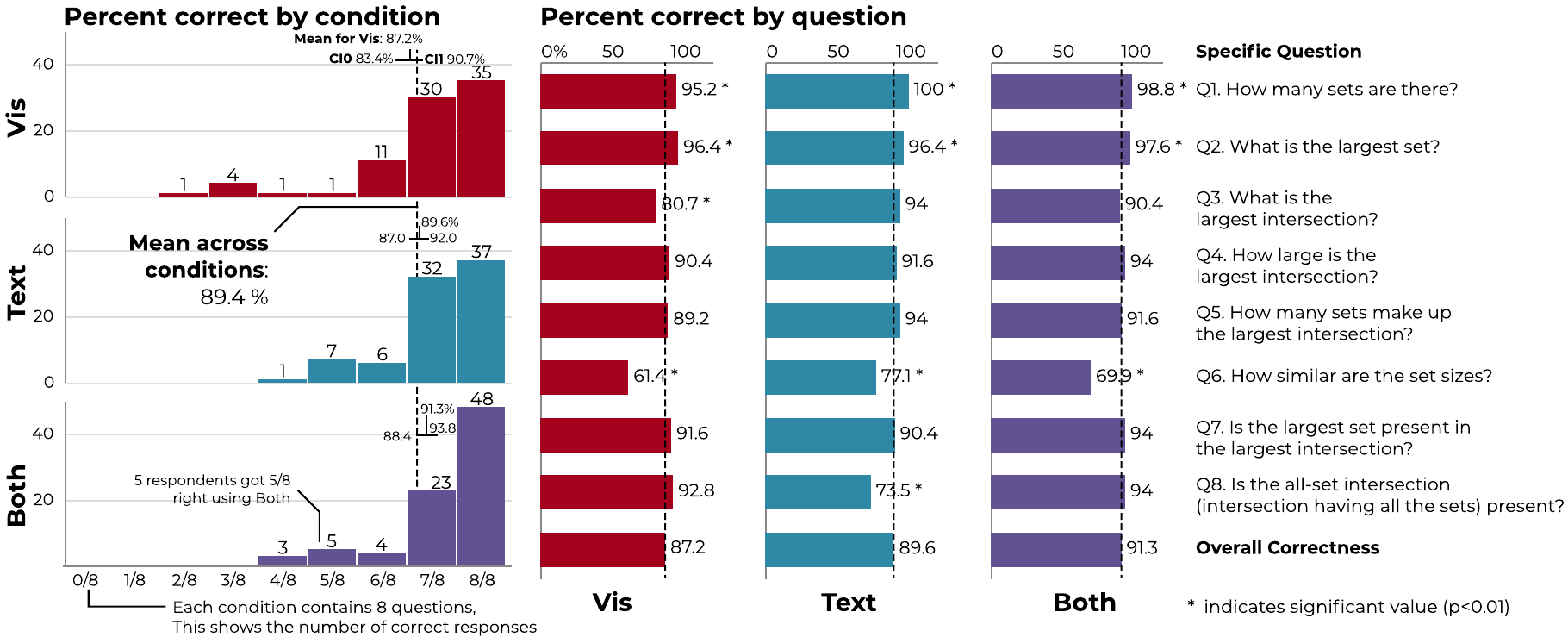}
\vspace{-3mm}
    \caption{
        Correct responses by condition (left) and by question (right).
        Each condition had \numSurvey{} participants, all of whom saw all  conditions.
    }
    \label{fig:correctness}
    \vspace{-2em}
\end{figure*}

\parahead{Alternative Presentations}
Two participants (\ix{2}, \ix{10}) with residual vision also viewed the COVID-19 UpSet plot with screen magnification.
Both participants reported that their mental model---a common practice for BLV users per Jung \etal{}~\cite{jung_communicating_2022}---created from the alt-text did not match the actual UpSet plot and specifically noted the matrix indicating the intersections (\cf{} \figref{fig:patterns}).
\ix{2} stated that, \qt{I read it more of just the size, like the section [with the bar graphs].}
Similarly, \ix{10} stated, \qt{I wasn't expecting the bubbles. The bar chart is what I was visualizing.}
Both participants stated that at their vision level, they would prefer visual UpSets to text because of the level of detail (\ix{10}) and convenience (\ix{2}).

Although most interviewees stated that the text description was sufficient at portraying the COVID-19 data, \ix{1}, \ix{3}, \ix{6}, \ix{7}, \ix{9}, \ix{10}, and \ix{11} expressed a preference for tactile displays.
We observed that participants compared tactile displays to the knowledge a sighted person would gain from a visualization.
As summarized by \ix{10}, \qt{if there is a tactile element associated with it, it'll be like someone looking at it. So it would be useful.}
\ix{6} also expressed similar association as they state \qt{I feel like that's the same, even though I can't see.}
Additionally, \ix{3} mentioned how a tactile display of an UpSet plot would help understand the layout of the visualization as \qt{it would be nice to see what an [UpSet plot] would look like in general.}
In future work we intend to explore how our automated captions can tie into physicalizations.

\section{Contextualizing Experiments}
\label{sec:experiments}

We complement our interview study with two experiments.
These studies are not key to our findings but offer context.
We elide most details (see appendix for them) and present only high level findings.

\subsection{Experiment 1: Utility for Sighted Users}

As a simplistic quality check, we considered if the affordances present in our text descriptions support the same level of analytic functionality as visual UpSet plots, and if our text descriptions can also be useful to sighted users, indicating a curb-cut effect where assistive infrastructure can benefit all users of that infrastructure, analogous to how video subtitles are useful for more than just language barriers~\cite{lundgard_accessible_2022, prasad_text_2012}.

To answer these question, we conducted a within-subjects crowd-work experiment with (N=\numSurvey{}) sighted users. Participants engaged with three conditions: just visualization (\condVis{}), just text (\condText{}), and text and visualization combined (\condBoth{}) across three datasets arranged in a Latin square.
For each dataset-condition combination we asked questions regarding each of the patterns described in \secref{sec:patterns}.
For \condVis{} and \condBoth{} we used static UpSet plots generated with our web tool.
For \condText{} and \condBoth{} we used the long form text-descriptions. In all conditions, we included a human-provided title and an introduction that leverages facts about the dataset (what are the items and sets). The experiment was implemented with reVISit~\cite{ding_revisit_2023} and is available \asLink{https://vdl.sci.utah.edu/Upset-alttxt-study/Upset-Alttext-User-Survey/}{online}.

While there are a variety of metrics of interest in this study, the key one is correctness: whether participants get answers right in each medium.
We hypothesized that there will be no statistical difference for correctness between the three conditions.
To support this question, we ran a mixed model. Correctness (reported as a percentage and summarized in \figref{fig:correctness}) was regressed onto all three conditions. Across conditions, approximately 6\% of the variance was explained by between-subject variability ($ICC$ = 0.06). Thus, we included a random intercept by participant to account for these inter-individual differences.
Correctness in the \condVis{} condition was significantly ($p$ = 0.01) less than in \condBoth{}.
However, correctness in text-only did not significantly differ from either other condition (\condBoth{} $p$ = 0.31; \condVis{} $p$ = 0.14).
While this rejects our hypothesis, the interpretation is straightforward: having two tools for analysis improves the results.
One participant explained \sqt{The combination of the text data and the visual graph is the easiest to understand in my opinion, providing most of the answers with a cursory glance but also providing the other answers with a bit deeper of a look.}
Suggesting that descriptions exhibit a curb-cut effect, making them useful for sighted people, as well as for a BLV audience.

The correctness parity between \condText{} and \condVis{} suggests that \textbf{our descriptions are, for sighted users, at least as good as the visual}.
However, we note that the ability to answer factual questions at similar correctness levels does not necessarily mean that \condText{} or \condVis{} confer benefits typically associated with visualizations---such as being memorable or getting understanding ``the gist'' of the data.
Finally, we note that this is the first experimental study of UpSet plot's usability and that it confirms that they are usable---as generally suggested from their widespread usage.

\subsection{Experiment 2: Description Replacability}

Here, we consider how well our bespoke descriptions improve over a naive replacement description.
As there has not been previously been work on captions for this chart form, there is not an existing baseline by which to compare. Instead, we use LLMs to generate text, the quality of which we explore through an ablation study (see appendix).
Across two studies we explore the effect of including different prompt components (such as the dataset, example descriptions, etc.), the impact of commonly known data for which context is available (like movies) vs anonymous data, and models (including OpenAI's GPT and Anthropic's Claude).
We then coded the descriptions using the semantic levels model~\cite{lundgard_accessible_2022} and conducted an expert review with our blind coauthor.

We find that while LLMs can produce high-quality text descriptions (sometimes in a more narrative format), they have nontrivial variance between generations that can contain false statements. The results are also dependent on the presence of related data in the models' training set.
While others~\cite{duarte_autovizua11y_2024} have previously used LLMs to successfully generate descriptive texts, we find these successes come with caveats.
We do not settle whether these risks are worthwhile, but find that our approach offers a consistent and reasonable quality compared to descriptions generated by LLMs.

\section{Discussion}
In this paper we explored how to develop accessible descriptive text for complex scientific charts in general by focusing on the specific case of UpSet plots.
We developed a system for automatically generating descriptions based on a survey of extant practices based on published usage. This system is general purpose and can be used to generate descriptive texts for any UpSet whose input can be summarized through a standard that we define.
Through this work we strive to make UpSet plots accessible to anyone.

\newcommand{\recipeWord}[1]{\textbf{\emph{#1}}}

\subsection{Transferring our Process to Other Chart Types}
Substantial study~\cite{tang_vistext_2023, nazemi_method_2013, obeid_chart-text_2020, engel_svgplott_2019} has already gone into understanding effective design of descriptions for basic chart types (such as line charts, bar charts, etc.), but there has been little consideration for more bespoke chart forms.
Outside of the UpSet specifics, a facet of our contribution is an example of a process for developing descriptions for complex or scientific chart forms.

We \recipeWord{synthesized} extant practices (via a survey of published charts). In particular, developing an understanding of not just the idioms of the chart type, but the patterns and practices that come from real usage were essential. These patterns and practices then can inform the choices of algorithms to surface them.

We found that it was essential to \recipeWord{develop language} around the data and layout patterns specific to UpSet plots.
Many charts enjoy a substantial body of extant patterns from which to draw upon in their descriptions, but new chart forms do not come ready-made with such descriptors.
Consideration of the patterns of and in chart types seems more broadly valuable, particularly as a component of inter-medium communication (\ie{} between visual and text).

We \recipeWord{co-designed} with a member of the community with whom we sought to work (via collaboration with a blind coauthor). This allowed us to catch many assumptions and not overfit to our biases.
We designed both our system and our descriptive text so they are \recipeWord{polymorphic}; that is they are not tied to a single usage context.

We \recipeWord{implemented} our approach in a platform-independent way, using a JSON standard and a web API, so that we do not solve the problem of text generation for a chart for a single implementation, but make it accessible to implementations in different languages.

We \recipeWord{validated} our approach in interviews with BLV users and provide templates for these interviews that can be adapted to validate text descriptions for other charts in the future.

We believe that this approach can be generalized to generate text descriptions for other non-standard chart types, such as network diagrams or tree maps.

\subsection{Limitations}
As with any work ours has a variety of limitations.
Our work centers on a single style of description. It is possible that other descriptions (such as by including the suggestions from our BLV interviewees) might have led to improved performance, however this was not examined. The goal of this work is not to perfect descriptions for UpSet plots (as such a goal is akin to writing the perfect greeting card), but instead to demonstrate they can be constructed in a consistent and useful manner.
Similarly, we focused only on a single chart type. If other types of set visualization had been considered as well, our results may have changed. Our focus was on UpSet specifically due to their relative popularity, but in the future comparing with other set visualizations (such as Euler diagrams) would be useful.

While we strove to incorporate real world UpSet practices, our text descriptions only covered a subset of all possible plots. For instance, our UpSet implementation includes advanced features (\eg{} aggregation, annotations) that are not currently captured by our descriptions. Future work should continue to add nuance to our descriptions to accommodate richer situations.

While we did ask some analytic questions of our interviewees, we did not perform a quantitative evaluation of BLV users' performance using our descriptions. Such studies would be valuable future work, but we emphasize that our goal was to demonstrate that they could be useful broadly, and, in doing so, we use these interviews as an opportunity to identify missing pieces in our design.

\subsection{Personalization}
Like any users, BLV users have varied preferences and expectations for presented information format (as Jones \etal{}~\cite{jones_customization_2024} note).
Some may prefer to have topics organized by bullet points, while others may prefer iteratively expanding summaries.
Similarly, this work exclusively focused on the accessibility of descriptions for BLV users, however other disabilities should be evaluated in the future---such as people with developmental disabilities~\cite{wu_understanding_2021}, like ADHD, or learning disabilities, like dyslexia or dysgraphia.
Moreover, additional work is likely necessary to explore how the visual representation of UpSet plots themselves might be adapted for people for whom understanding multiview visualizations is challenging.
Echoing Wu \etals{}~\cite{wu_our_2024} findings that narrativization is useful for people with intellectual disabilities, we suggest this type of affordance is useful for any group. This introduces the possibility of false-information being induced to the output, which we should design around carefully.

\subsection{Interactivity}
While the web tool we built supports \textbf{interactive analysis} (including for BLV users by regenerating the descriptions on specification change), the descriptions do little to support interactivity directly. Future work might explore how user interaction might be woven into these descriptions to support more dynamic analysis. Alternatively, an approach similar to Zong et al's rich screen reader experiences \cite{zong_rich_2022}, where BLV users can navigate the primary chart, e.g., using data navigator~\cite{elavsky_data_2024} is conceivable.

\section{Acknowledgments}

We gratefully acknowledge funding by Chan Zuckerberg Initiative Essential Open Source Software for Science program.

\bibliographystyle{eg-alpha-doi}

\bibliography{2024-upset-alttext}

\clearpage
\onecolumn

\appendix

\section{Appendix Introduction}

In this appendix we include, as appendices, various details and features which did not succinctly fit in the main text.
First, in \secref{sec:resources} we collect all links to supplementary resources on the web. In \secref{sec:survey}, we provide a full description of our contextualizing experiment in which we surveyed sighted users about their ability to user our text descriptions.
Then, in \secref{sec:llms}, we provide a full description of our study of LLM's ability to produce alternative texts.
Next, \secref{sec:instruments}, we provide the instruments used in our various studies.
We then, in \secref{sec:text-descriptions}, provide examples of text descriptions.

\section{Resources}
\label{sec:resources}

\noindent The OSF repository containing all supplementary material: \\
\url{https://osf.io/kbvs9/}

\noindent The UpSet text description generation code: \\
\url{https://github.com/visdesignlab/upset-alt-txt-gen}

\noindent  The live-version of the web-based UpSet tool with text descriptions: \\
\url{https://upset.multinet.app/}

\noindent The code for the web-based UpSet tool: \\
\url{https://github.com/visdesignlab/upset2}

\noindent The modified version of UpSetPlot that can be used to generate text-descriptions via Python API call: \\
\url{https://github.com/JakeWags/UpSetPlot/tree/alt-text-generation}

\noindent An example notebook illustrating how to generate text descriptions in Python:
\\
\url{https://github.com/visdesignlab/UpSet-Survey-Data-Analysis/blob/main/suppl_upsetplot.ipynb}

\subsection{Interview Study}

\noindent The code for the stimuli used in the interview study: \\
\url{https://github.com/visdesignlab/upset-alttext-example}

\noindent The deployed version of the stimuli: \\
\url{https://vdl.sci.utah.edu/upset-alttext-example/}

\noindent \subsection{Crowdsourced Experiment}
The code defining the crowdsourced experiment: \\
\url{https://github.com/visdesignlab/Upset-alttxt-study}

\noindent The deployed crowdsourced experiment: \\
\url{https://vdl.sci.utah.edu/Upset-alttxt-study/Upset-Alttext-User-Survey/}

\noindent The code for the data analysis: \\
\url{https://github.com/visdesignlab/UpSet-Survey-Data-Analysis}

\section{Contextualizing Experiment 1: For sighted users, are our descriptions equivalent to the UpSet plots? }
\label{sec:survey}

As simplistic quality check, we sought to understand if the affordances present in our text descriptions are equivalent those in the visual representation of upset plots.
To answer this question we conducted a within-subjects crowd-work experiment with sighted users. Participants engaged with three conditions: just visualization (\condVis{}), just text (\condText{}), and text and visualization combined (\condBoth{}).
We included the \condBoth{} condition to see whether there are benefits of including text descriptions even when visualizations are present.
Through this experiment we sought to answer the following questions:

\newcommand{\hypo}[1]{$H_{\text{#1}}$}
\newcommand{\hAcc}{\hypo{accuracy}}
\newcommand{\hPref}{\hypo{preference}}
\newcommand{\hTime}{\hypo{time}}
\newcommand{\hVibes}{\hypo{insights}}

\begin{description}
    \item[\hAcc{}:]
          \emph{Can participants answer factual questions about UpSet plots equally well in the three conditions?}
          We hypothesize that there will be no statistical difference for correctness between the three conditions.

    \item[\hTime{}:]
          \emph{Does the description-only condition take significantly longer to read than the visualization conditions?}
          While time is not an essential attribute for most analysis tasks (if the differences are small), we speculate that the \condText{} condition will lead to slower response times, because text has to be scanned linearly.

    \item[\hPref{}:]
          \emph{Which conditions will participants prefer?}
          We speculate participants will prefer the \condVis{} and \condBoth{} conditions and dislike the just \condText{} condition. We expect measures of confidence, effectiveness and understandability to be lowest for \condText{}.

    \item[\hVibes{}:]
          \emph{Are the types of insights that are available through only \condText{} substantially different?}
          We speculate that \condVis{}/\condBoth{} lead to ``higher level'' synthesis (\eg{}  trends or correlations) and that \condText{} leads to re-stating facts.
\end{description}

With respect to our top-level question, we find that performance with \condText{} and \condVis{} are roughly equivalent in relevant dimensions. In addition we find that \condBoth{} tends to perform better, suggesting that descriptions (or at least simplified alternative presentations) have value for usages other than simply making charts more accessible.

\subsection{Procedure}

The experiment involved four principal phases: consent, tutorial, experiment, and post-experiment survey.
This study was reviewed by our institution's IRB and deemed exempt from full board review.
We used reVISit~\cite{ding_revisit_2023} to develop and deploy the study.

\parahead{Participants and Pilots}
We recruited N=\numSurvey{} participants from the US (21), UK (30), and Canada (32) using Prolific.
Subjects were eligible to participate if they were fluent in English and over 18.
Participants took about 27 minutes on average and were compensated \$8 USD, for an average hourly pay of about \$17 USD.
They were on average $30.7 \pm 9.6$ years old. 31 Female/ 52 Male.

We conducted down-the-hall pilot with (N=4) users and then a full pilot on (N=5) Prolific users.
These revealed that users who did not read the introduction did poorly in the experiment.
In response, we required that the training questions be answered correctly after three tries.
This requirement may have led to a high return rate: 84 participants completed the study, 8 were rejected because they did not complete the study, 80 returned the study after reading the warning.
During analysis, we discovered one participant was not completely logged in the raw file. The participant completed the training and first condition (\condBoth{}), but the answers to the rest of the questions were missing and so we excluded them.
to be consistent with the rest of the study.

The experiment began with an elicitation of consent and an attention check tutorial.
For each dataset we asked questions regarding each of the patterns described in \secref{sec:patterns}---for instance, whether or not the all-set intersection is present. We list these questions in \autoref{fig:correctness}.
We did not explicitly ask about intersection patterns, such as the presence and prominence of high-order set intersections because these responses require higher-level synthesis and are not easily answered succinctly. Instead, we used open-ended questions about the insights and takeaways.
In addition, we asked for participants' confidence in their answers, their understanding of the information, and effectiveness of the presentation on a 1-5 Likert Scale (5 being high).

\begin{figure}[t]
    \centering
\includegraphics[width=\linewidth]{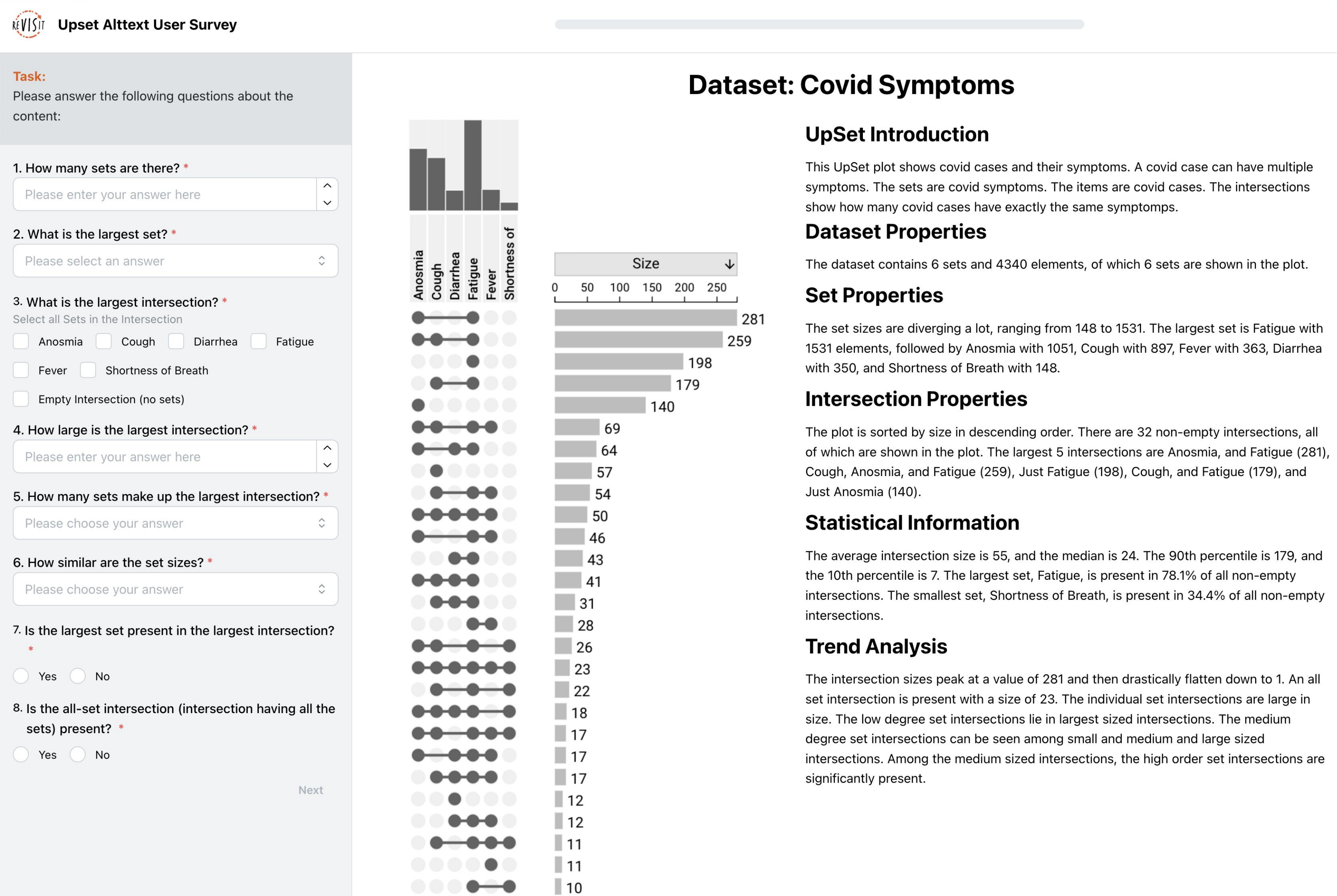}
\caption{Example of the study interface for the \condBoth{} condition. The title (right) is human-generated, the introduction leverages a human-provided specification. See \linkToStudy{} for more.
    }
    \label{fig:study-interface}
\end{figure}

\autoref{fig:study-interface} shows an example of the study in the \condBoth{} condition.
For the \condVis{} and \condBoth{} condition we used static UpSet plots generated with our web tool and sorted by size.
For the \condText{} and \condBoth{} conditions, we used the text-description as provided by the long form. In all conditions, we included a human-provided title and an introduction that leverages human-provided facts about the dataset (what are the items, what are the sets).

We used four different datasets:
\begin{description}
    \item[Movie Genres.] Movies and movie genres~\cite{harper_movielens_2016}. 6 sets. Intersections are low-degree. Used only for the training stage.
\item[Tennis.] Tennis players who won one or multiple of the four grand slam tournaments ~\cite{west_upsetr}. 4 sets. Mix of single-set intersections and high-degree intersections. All-set is prominent. See \autoref{fig:upset_explained}.
    \item[Organizations.] Countries and their participation in global organizations (\eg{} the WHO or NATO)~\cite{centralintelligenceauthority_international}. 8 sets. Very unbalanced set sizes, lots of high-degree intersections, but no all-set intersection. See \autoref{fig:patterns} right.
    \item[Covid.] COVID-19 symptoms~\cite{lyi_altair-based}. 6 sets. Unbalanced set size, many combinations of symptoms present, including all-set intersection.
\end{description}

To counterbalance learning effects, we used a Latin square to permute condition assignment (\condVis{}, \condText{}, \condBoth{}) to datasets (Tennis, Organizations, Covid), and to permute the order in which the conditions would appear in the sequence (yielding a 3x3x3 square).
The high return rate led to a slight imbalance in dataset-condition combinations.
On average, 27.6 user saw each combination.
We believe that these minor variations do not affect our results.

\parahead{Post-Experiment Survey}
The session concluded with a short survey covering preferences, perceived effectiveness, readability and description length (rated on a 1-5 Likert scale, 5 high), as well as open-ended general comments.

\subsection{Limitations}

The high return rate of our study led to a slight in-balance between groups in the Latin square, and data for one participant was lost, possibly due to connectivity issues. While unfortunate, we believe the effect of these errors to be minor and to substantially not effect the content of our analysis.
The design of our survey focused on simple analytic tasks, instead of more complex exploratory-type tasks. While investigating these areas would be useful, the goal of this chart in publications is typically presentational rather than exploratory.

\subsection{Results}
\label{sec:quantresults}

Next we describe the results of this study incorporating both the quantitative results, as well as analysis of the free text results.
We refer to our Survey Participants as \sx{X} and \sqt{quote them}.

\begin{figure}
    \centering
    \includegraphics[width=0.8\linewidth]{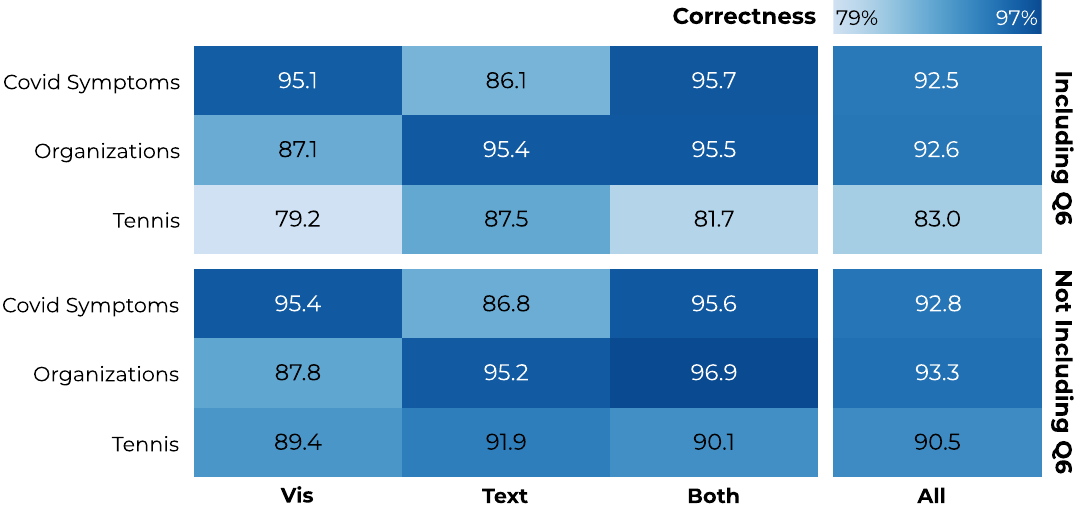}
\caption{Percent of questions correctness by dataset and condition. Q6, which focuses on qualitative distribution similarity, had a high influence on the average correctness by dataset.
    }
    \label{fig:dataset-correctness}
\end{figure}

\parahead{Correctness}
To evaluate whether participants were more correct in one condition over the others (\hAcc{}), we ran a mixed model. Correctness (reported as a percentage) was regressed onto all three conditions. Across conditions, approximately 6\% of the variance was explained by between-subject variability ($ICC$ = 0.06). Thus, we included a random intercept by participant to account for these inter-individual differences.
These correctness results are summarized in \autoref{fig:correctness}.Correctness in the \condVis{} condition was significantly ($p$ = 0.01, $\text{BF}_{\text{01}}$ = 0.86 less than in \condBoth{}.
As frequentist statistics cannot provide evidence in favor of the null hypothesis, we also report Bayes Factors ($\text{BF}_{\text{01}}$). $\text{BF}_{\text{01}}$ assesses how strongly the evidence supports the null hypothesis ($H_{0}$) compared to the alternative hypothesis ($H_{1}$). A $\text{BF}_{\text{01}}$ greater than 3 is considered strong evidence for the null hypothesis and A $\text{BF}_{\text{01}}$ less than 0.3 is considered strong evidence for the alternative hypothesis.
However, correctness in text-only did not significantly differ from either other condition (\condBoth{} $p$ = 0.31, $\text{BF}_{\text{01}}$ = 12.62; \condVis{} $p$ = 0.14, $\text{BF}_{\text{01}}$ = 8.33).
While this rejects \hAcc{} the interpretation is relatively straightforward: having two different tools for analysis improves the results.
\sx{29} explained the benefits of using the two modes in tandem: \sqt{The combination of the text data and the visual graph is the most easy to understand in my opinion providing most of the answers with a cursory glance but also providing the other answers with a bit deeper of a look.}
This seems to suggest that having descriptions is useful for sighted people as well as for their expected BLV audience (which we explore in \secref{sec:interview}).
Following prior works~\cite{lundgard_accessible_2022, prasad_text_2012}, we suggest that this is a curb-cut analogous to how video subtitles are useful for more than just those with a language barrier.

More closely related to the core of our study, the equivalence of \condText{} and \condVis{} for this measure suggests that \textbf{our descriptions are, generally, at least as good as the visual} which is a key result for this study.

Next, to examine whether the correctness across conditions were driven by a particular question, we added the eight questions (Q1-Q8) to the mixed regression. The questions were effect coded so that the regression coefficients reflect comparisons to the overall average correctness under each condition. A likelihood ratio test reveals that adding the questions as a predictor to the model significantly improves the model's fit ($\chi^2(21)$ = 207.48, $p <$ 0.01).
This highlights that ability with a given condition is \emph{not} linked to the type of question being asked, and is instead based on the difficulty of the question itself.
This again supports a conclusion that \condText{} and \condVis{} are functionally equivalent.

We evaluated several different factors to verify our results.
Self-reported visualization \textbf{expertise} (using a 1-5 Likert scale) and correctness were uncorrelated ($p$=0.208), suggesting that this finding is broadly applicable.
\textbf{Presentation} order did not significantly effect correctness.
Via a regression analysis, we found that across all conditions, correctness improved by ~1\% on average from first block to second block ($p$=0.55, $\text{BF}_{\text{01}}$ = 19.68) and $\sim2\%$ from second block to third ($p$=0.27, $\text{BF}_{\text{01}}$=11.02).

Finally, we found that \textbf{dataset} had substantial effect on correctness. In \autoref{fig:dataset-correctness} we see that participants performed significantly worse with the Tennis dataset  across all conditions, and especially so in the conditions that showed a visualization, compared to the other datasets.
This difference is caused mainly by Q6 (How similar are the set sizes) which had answers options ``roughly equal'', ``diverging a bit'', ``diverging substantially'' (see also Q6 in \autoref{fig:correctness}).
The tennis dataset is unique with respect to this question, because the answer for the size of sets is different from the answer for the size of the intersections.
In contrast, for the other datasets the answer is (coincidentally) the same.
We speculate that some participants mistakenly answered for intersections when answering based on the plot, which led to a larger number of wrong answers.
This would seem to suggest a limitation of UpSet plots. As both sets and intersections can be interpreted as being on an equal visual footing, it can be easy to confuse them. At the same time, this potential for confusion is not present in the text description, likely leading to better responses.
Overall, this suggests that UpSet plots are more effective in cases where that ambiguity is not relevant to the intended message or analyses.

\begin{figure}[ht]
    \centering
    \includegraphics[width=0.6\linewidth]{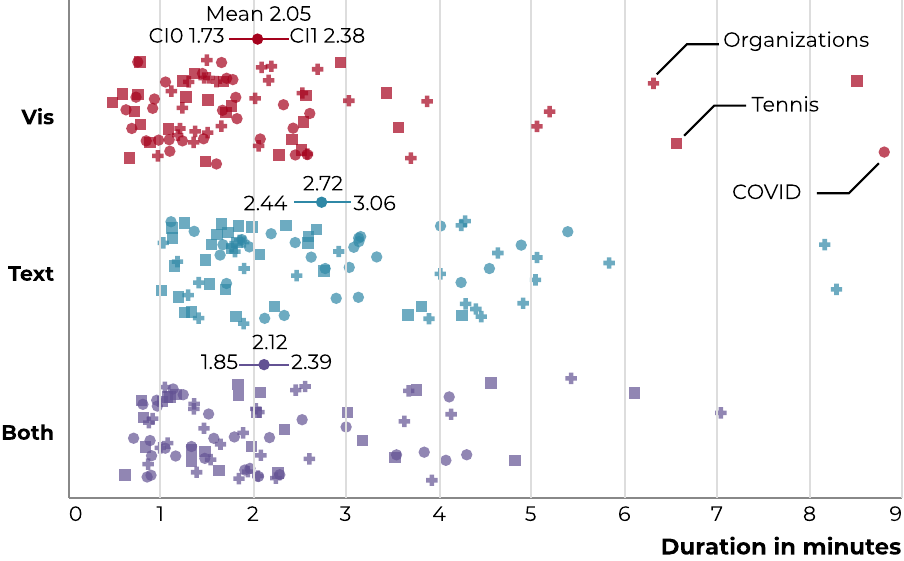}
\caption{
        Time to completion by section and condition. Participants take longer to respond in the text condition.
    }
    \label{c}
\end{figure}

\begin{figure}[!htb]
    \centering
    \begin{minipage}{.5\textwidth}
        \centering
        \includegraphics[width=\linewidth]{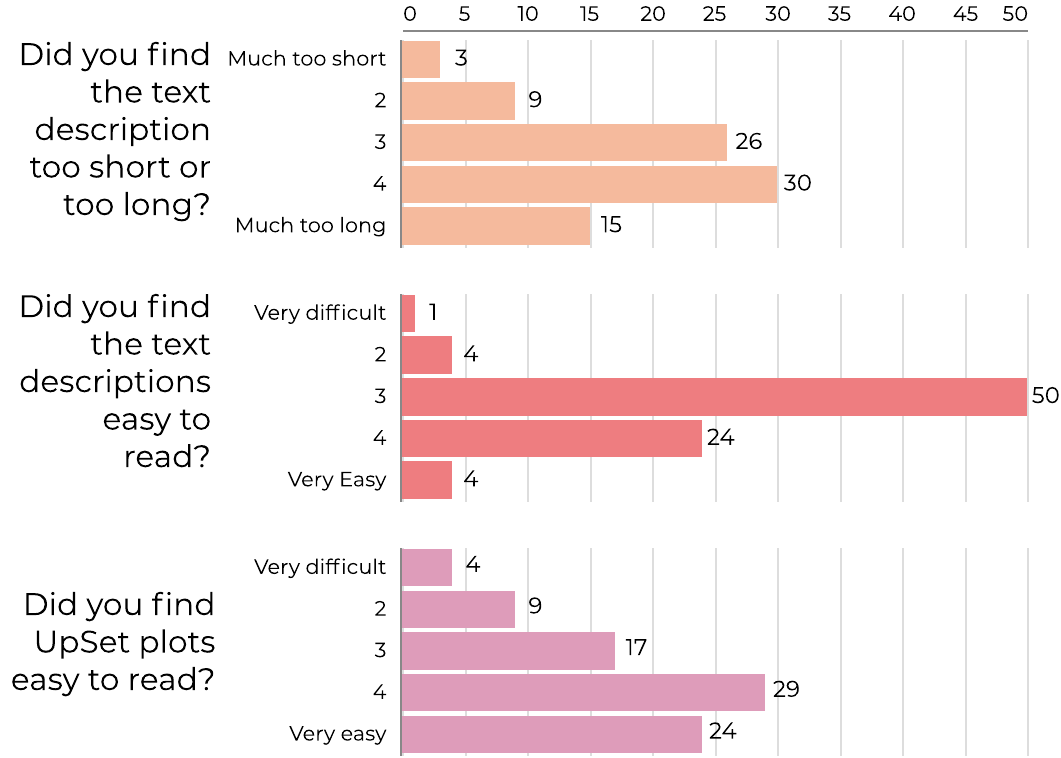}
        \caption{Espoused preferences from survey respondents in the post-experiment survey. }
        \label{fig:survey-prefs}
    \end{minipage}\begin{minipage}{0.5\textwidth}
        \centering
        \includegraphics[width=\linewidth]{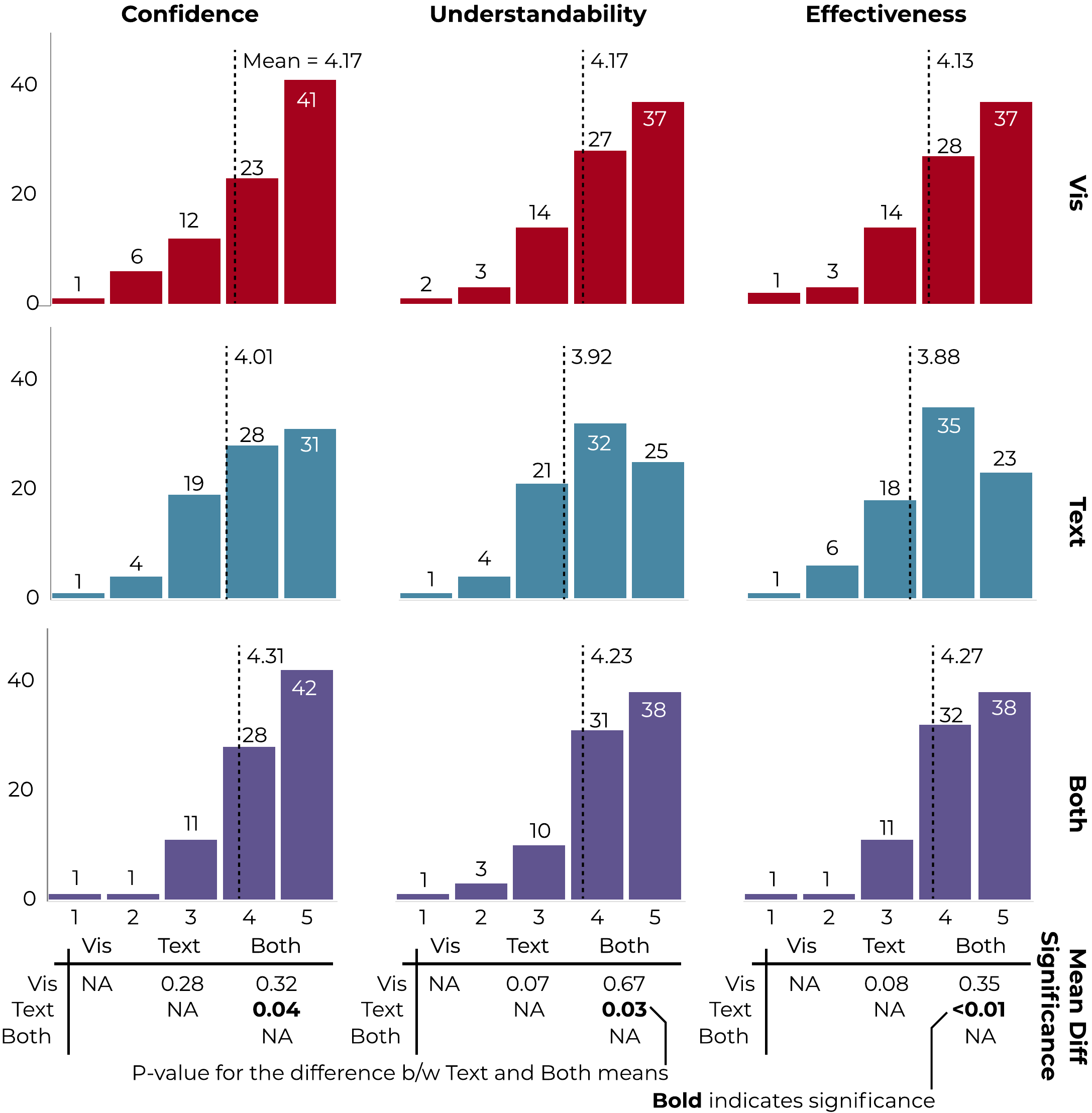}
        \caption{Survey respondent reported confidence, understandability, and effectiveness for each of the three conditions via 1-5 Likert items (5 being high). Top rows show histogram by count, bottom shows significance of means. }
        \label{fig:confidence}
    \end{minipage}
\end{figure}

\parahead{Completion Time}
Next we consider the effect of condition on completion time.
We again use a mixed effects model in which completion time (in minutes) was regressed onto the three conditions, with a random intercept by participant to account for high between-subject variability ($ICC$ = 0.60).
Per \autoref{fig:confidence}, the average completion time was about 2 minutes across conditions. \condText{} took slightly longer---averaging 2 minutes 45 seconds vs \condVis{}'s 2 minutes 3 seconds---which was a significant divergence from the other conditions (\condText{} vs. \condVis{} $p <$ 0.01, $\text{BF}_{\text{01}} <$ 0.01; \condText{} vs. \condBoth{} $p <$ 0.01, $\text{BF}_{\text{01}} <$ 0.01).
The other conditions did not significantly differ (\condVis{} vs. \condBoth{} $p =$ 0.34, $\text{BF}_{\text{01}}$ = 16.82).
\sx{55} summarized that the text was \sqt{easy to understand and read although missing the visual component takes away from the speed you are able to understand the information.}
Similarly, \sx{13} noted \sqt{Taking a little more time to crosscheck visualization with text gives greater (confidence in) accuracy and it is still very fast to parse the data.}
These results broadly agree with our hypothesis (\hTime{}), however taking an additional 40 average seconds does not seem like an unduly high burden for a more linear medium.
Prasad \etal{}~\cite{prasad_text_2012} find for simple datasets and charts, text leads to more correct answers, but slower responses than charts or tables. Our results generally agree with this finding.

\parahead{Preferences}
Next, we consider participants' espoused opinions about each of the conditions. For participants ratings of confidence, understandability, and effectiveness we regress onto the three conditions (\autoref{fig:confidence}).
We find that the only significant differences among these preferences are, that for all three measures, the \condBoth{} condition is preferred over text---echoing our correctness findings.
Similarly, participants strongly preferred \condBoth{} over the other two; 57/83 participants preferred \condBoth{}.
\sx{49} commented \sqt{Having both text description and data visualization combined works the best since the people who can read UpSet plots will use the plot to extract information, but the people who do not understand UpSet plots [...] can use the text description to help them get information about the dataset.}
See also \autoref{fig:survey-prefs}.

Connecting preference with correctness, we suggest that the lower on average correctness for \condVis{} can be explained by these charts being unfamiliar, leading to a lack of confidence in their answers (compared to \condBoth{}).
For instance, \sx{20} noted \sqt{As this is a new way of interpreting data, I am not as confident without the written explanation alongside to support my understanding.}
Participants did not find UpSets to be easy (nor hard) to read, rating them $3.72 \pm 1.14$ out 5, with 5 being ``very easy''.
\sx{51} explained that \sqt{some things were simple to see, such as number of sets and intersections. However I am not generally confident relying on just visualization. Although this may be because this method of displaying data is new to me.}
This, again, highlights the value of descriptions for all, and not just BLV users.
These results partially reject \hPref{} as participants seem relatively indifferent between \condText{} and \condVis{} compared to their preference for \condBoth{}.
As with correctness, this preference seems natural as \condBoth{} offers best-of-both affordances.

\begin{figure}[h]
    \centering
    \includegraphics[width=0.3\linewidth]{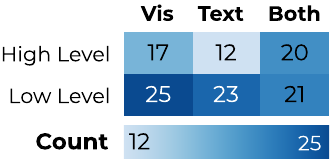}
\caption{Insight found condition and type of insight.}
    \label{fig:insight-count}
\end{figure}

\parahead{Types of Data Insights}
Next we consider if condition effected the types of insights (here meaning broadly facts about the data) that participants had.
We used a deductive coding scheme~\cite{king_template_1998} to code the 348 non-blank free-text responses.
We met iteratively to understand the codes. We focus on those tagged \emph{high level insight} or \emph{low level insight}.
High-level covers aspects related to the dataset. For instance \sx{43} observed, \sqt{Higher probability of being a member for Interpol, UN, UNESCO, UPU and WHO. Nobody has been a member for all organizations} concerning the Organization dataset.
Low-level includes elements like counts or basic comparisons, such as in \sx{53}'s note that \sqt{fatigue is the most common symptom} for the Covid dataset.
We count these insights by condition in \figref{fig:insight-count}.
This is a relatively low sample size, and many observations in the set were not data related. Yet, it appears from this sample that lower-level commentary is more frequent, and that the \condBoth{} had 1.6x more high-level comments than \condText{}. This rejects \hVibes{} as the types of insights appear to be roughly equivalent between \condText{} and \condVis{}.

\parahead{Feedback}
While participants generally found the descriptions to be informative, there was still room for improvement.
\sx{9} observed that \sqt{I think the text description could benefit from more formatting. It's difficult to parse the long sentences.}
Similarly, \sx{14} suggested that \sqt{the text description were so procedural in their style that it was difficult to tell what information it was trying to give me with the data. It seemed with that much text that there would be more explanation of the actual nature of the data.}
Not everyone felt it was hard to parse \condText{}. For instance, \sx{44} commented that \sqt{the headings are split up nicely, so finding the section you want to look at is quite easy.}
\sx{30} observed that there was a slight learning curve to the text, noting \sqt{after understanding how to read the information, I found it very simple to find exactly what I needed in a short amount of time to answer the questions}.
Similarly, on average participants were neutral about the readability of the descriptions rating $3.54  \pm 1.02$ out of 5, as well as its length (averaging $3.31 \pm 0.70$ with 5 being too long and 1 being too short).
This diversity of opinions highlights that description personalization may be valuable: what is useful for one person may not work for another (echoing Jones \etal{}~\cite{jones_customization_2024}).
\sx{66} emphasized this point by noting how reformatting could improve accessibility generally: \sqt{I think overall maybe having less text and more bullet points would make it easier and quicker to get the gist of what's being depicted (especially for folks who have ADHD or other learning disabilities like myself.)}
We note that evaluating description usability for people with other kinds of disabilities is valuable future work.

\newcommand{\colabName}{{Dr. X}}
\newcommand{\ablationRun}[1]{\textbf{AB}$_{\textit{#1}}$}
\newcommand{\bestEffortRun}[1]{\textbf{BE}$_{\textit{#1}}$}

\section{Contextualizing Experiment 2: Are our generations better than a default replacement?}
\label{sec:llms}

Here we consider how well our bespoke descriptions improve over a naive replacement description.
As there has not been previously been work on captions for this chart form, there is not a existing baseline by which to compare. Instead, we use LLMs to generate text, the quality of which we explore through an ablation study.

We find that while LLMs can produce high-quality text descriptions (sometimes of a more narrative format), they have non-trivial variance between generations (yielding inconsistent quality) that can contain false statements. The results are also dependent on the presence of related data in the models' training set.
While others~\cite{duarte_autovizua11y_2024} have previously used LLMs to successfully generate descriptive texts, we find these successes come with caveats.
We do not settle whether these risks are worthwhile, but find that our approach offers a consistent and reasonable quality compared to descriptions generated by LLMs.

\subsection{Procedure}

To answer the question how LLMs compare to our approach, we ran two comparison studies.
First we ran an ablation study in which we explored the effect of several components of the generation---but kept the model and dataset fixed.
We refer to runs in this study by \ablationRun{X}.
Second, we ran a ``best effort'' study where we varied model and dataset with the best performing combination of factors from the ablation study.
We refer to runs in this study by \bestEffortRun{X}.
All studies were performed in late summer 2024.
For generations the model temperature was set to zero.

\begin{figure}
    \centering
    \includegraphics[width=\linewidth]{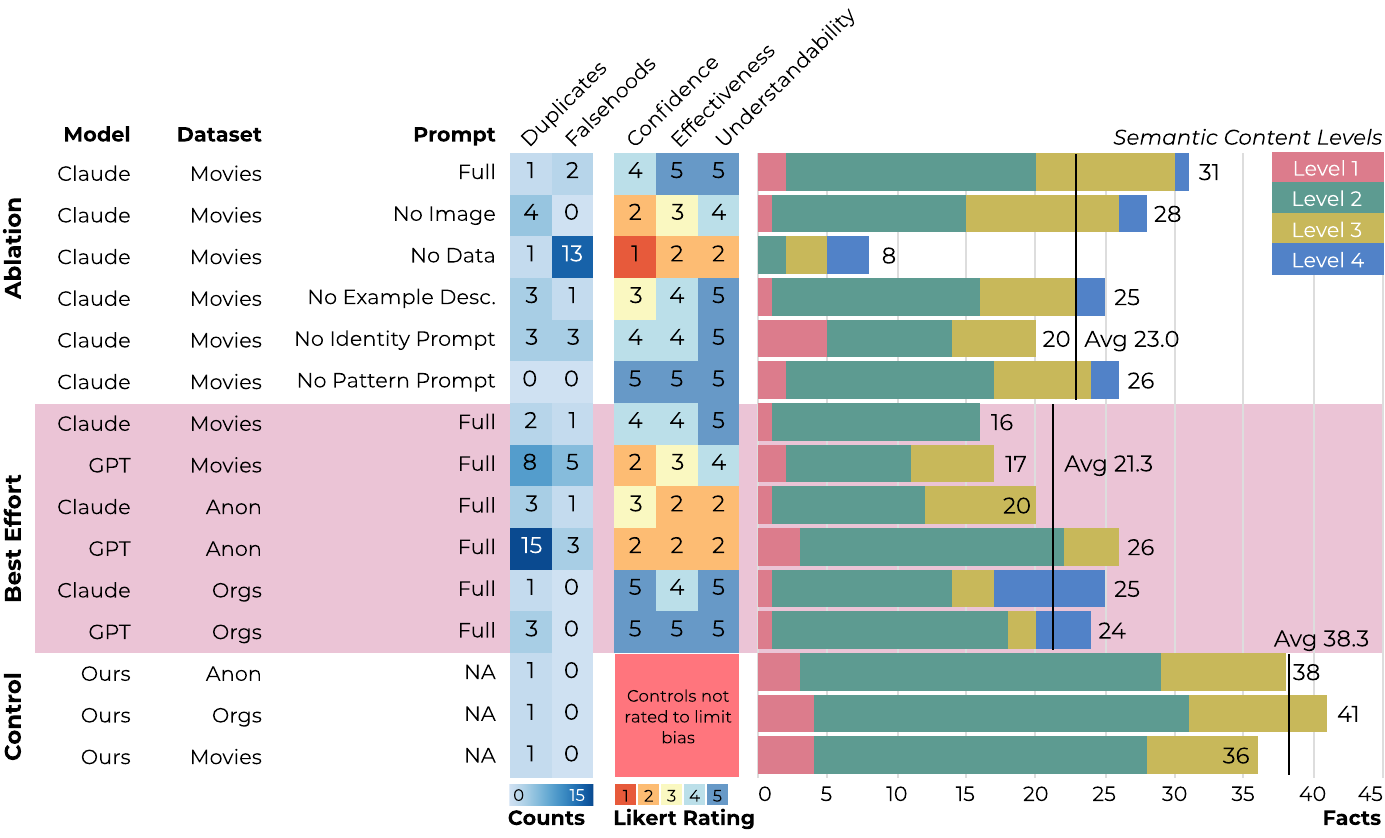}
\caption{A summary of our LLM study showing study instance by the number of facts generated for that description, sorted by Lundgard \etals{}~\cite{lundgard_accessible_2022} semantic content levels. We also highlight the number of hallucinated and duplicated facts in the generation---although, as they were identified manually, these values are lower bounds, as there might be additional subtleties to the data not analyzed. Claude here is Anthropic's Claude 3.5 Sonnet and GPT is OpenAI's GPT-4o. N.b. the inputs for the Best Effort Movies  run and the Ablation Claude Movies were identical.}
    \label{fig:llm-study}
    \vspace{-2em}
\end{figure}

\parahead{Setup} For the \textbf{Ablation study} we ablated across the following components:
\emph{Image} (including an image of the UpSet plot in question),
\emph{Accessible Processed Data} (including the data being analyzed),
\emph{Example Descriptive Text} (including an example text description we created),
\emph{Strong prompt} (includes both a role play prompt focused on the creator of UpSet plots as well as guidance on effective text generation), and
\emph{Pattern prompt} (prompt summarizing the data patterns from \secref{sec:patterns}).
For the \textbf{Best} Effort study, we used an ``all on'' prompt strategy.
We vary models between Claude 3.5 Sonnet and OpenAI's GPT-4o. These models were selected because they offer good performance~\cite{kane_claude_2024}, are widely available, and can incorporate both text and images (within their respective ecosystems). See \autoref{fig:llm-study} for model use by run.
We vary datasets between Organizations, Movies, and an anonymized version of Movies called Anon-Movies  (\ie{} the same data with set names replaced with ``set\_1'' and so forth).
We are interested in the difference between Organizations and Movies because they have very different data distributions (as described in \autoref{fig:patterns}), as well as the difference between Anon-Movies and Movies for the effect of context that can be inferred.

\parahead{Evaluation} We took two approaches to comparing the LLM generated text and with ours: coding and expert review.
First, we had two coders independently code the result of the ablation study.
They counted the number of correct and incorrect facts in each generation and reviewed the content present using Lundgard \etals{}~\cite{lundgard_accessible_2022} semantic levels framework.
After coding they met to discuss and resolve ambiguities, after which they adjusted their original codes.
With this more robust coding model in hand the first coder then coded the best effort study independently.
An example of a coding of one our generated text is shown in \figref{fig:set-example}.
We do not code for our identified data patterns because they are part of some prompts.
Next, we had our blind collaborator, \colabName{}, evaluate the LLM generated responses across the same metrics used in \figref{fig:confidence}, as well as offer any qualitative commentary he had.
We quote him \eqt{like so}.

\parahead{Limitations}
Our evaluation of the LLM generated descriptions via grading may not reflect the way in which real users perform with them. This analysis was primarily centered on demonstrating that our results are not trivially replaced by an LLM call rather than the comparative effectiveness---but having users evaluate these properties is interesting future work.
Our grading of the generated descriptions may have been biased. We tried to limit this effect by having two coders develop a code book, however these efforts will inevitably be incomplete.
Similarly our blind collaborator's rating of the generated texts may be biased, but given the range of his ratings we feel this is unlikely.
Finally, the quality of the LLM results could be improved through additional prompt engineering. The goal of this evaluation was to see if a simple or straightforward usage would be better than our handcrafted descriptions, but integrating LLMs more deeply into text generation is intriguing future work.

\subsection{Results}

Here we explore the results of our LLM study.
\autoref{fig:llm-study} shows a summary of our results.

\parahead{Text Content Differences}
On average our descriptive text contained a substantially greater number of facts.
Interestingly, our texts tended to be shorter (avg 290 words), compared to the ablation runs (avg 325) and best effort runs (avg 384)---this is unsurprising given the tendency for LLMs to be verbose.
However, measuring quality through text length is noisy as LLM responses can be made more verbose or concise via prompt engineering (of which we did little, beyond the coarse component-level exploration in our ablation study).
Compared to the texts we designed together, \colabName{} liked \eqt{metaphorical language, such as a `long tail distribution'} that was present in some LLM texts, noting of \ablationRun{No Pattern} that \eqt{It's the kind of alt text I'd expect to see on a news website.}

However, some runs also had  substantial numbers of falsehoods, some of them were obvious (and caught in our coding) while others of them might be subtler. For instance \colabName{} worried that the LLM had \eqt{cherry picked certain interesting elements it wanted to highlight} or \eqt{misinterpreted} certain data. Overall GPT seemed to do worse in this regard, as it more frequently generated duplicate and false information---although it could be tuned to perform better.

Finally, the generated texts have some variance. \ablationRun{Full}  and \bestEffortRun{Claude-Movies} were relatively different (\eg{} the volume of Level 4 content) but had the same inputs.
If something is to be relied upon for accessibility, its value may be limited if it changes day to day.

\parahead{Knowledge is Key}
The presence of data and that data being familiar seems to be an essential part of making effective LLM texts.
For instance, \ablationRun{No data} generated a substantially lower number of facts and was qualitatively worse for \colabName{}: \eqt{I can't build a mental picture of what the visualization might look like, or what story the data tells me.}
This suggests that including the dataset is necessary for providing accurate information.
In addition, that data seems to need to have human-readable labels.
While there was relatively little difference between the types of semantic content generated in the Anon Movies and Movies runs, the effect of reading those texts was starkly different.
\colabName{} observed that \bestEffortRun{*-Anon Movies} were \eqt{hard to follow} and \eqt{hard work to get something sensible out of it.}

The LLM generated texts include some amount of Level 4 semantic content, which we intentionally excluded from our design.
Notably, the \bestEffortRun{*-Orgs} runs included substantially more Level 4 content than compared to those using other data.
For example, \bestEffortRun{Claude-Orgs} included observations like \texttt{``This underscores the tendency for nations to participate in multiple international bodies rather than isolating themselves to a single organization.''}
This is likely because news articles and other information about organizations like the WHO are present in their training data.
\colabName{} specifically liked the amount of level 4 content in the Orgs runs because it provided specific examples to back up claims.
Evidently, if the model being used has knowledge of the dataset under study, then it can usefully provide contextualizing information. However, this can be risky as it is not clear when a model will or will not know about a dataset.

\section{Study Instruments}
\label{sec:instruments}

Here we show simplified versions of the study instruments used in the survey and interview studies

\subsection{Survey Study}

An anonymized version of our survey is available at \linkToStudy{}. This includes a full copy of the instructions participants were shown and supports browsing each of the study conditions.

\subsection{Interview Study}

\subsubsection{Demographic Information}
\begin{enumerate}
    \item What is your date of birth?
    \item What is your gender?
    \item What is the highest level of education you have completed or are in the process of completing?
    \item Which screen reader do you use with your computer or other devices (e.g., NVDA, JAWS, VoiceOver, etc.)
    \item How long have you been using a screen reader?
    \item Do you use other accessibility devices or software in combination with a screen reader, such as screen magnification or a braille display? If yes, please describe.
    \item What is your preferred rate of speech when using a screen reader?
    \item How many hours are you on a computer each day?
    \item Would you consider your career to be data-intensive or numbers-driven (e.g., regularly work with large datasets, perform statistical analyses, or make decisions based on quantitative information)?
    \item How often do you interact with data visualizations, such as those for work, from news articles, in video games, etc.? And in what context?
    \item How would you describe your vision-loss level?
    \item How would you describe your vision level (e.g., no remaining vision, light perception, central vision, etc.)?
    \item What is your corrected visual acuity in either Snellen (e.g., 20/200) or LogMAR (e.g., 1.3)?
\end{enumerate}

\subsubsection{UpSet Alt-Text Questions}
\begin{enumerate}
    \item Can you describe what you learned about the dataset in your own words?
    \item What is the dataset about?
    \item How many sets are shown?
    \item What is the largest intersection?
    \item How would you describe this dataset to a friend?
    \item How did this dataset increase your understanding of COVID symptoms?
    \item Do you feel like you have a good sense of the dataset? Why, why not?
    \item What was difficult for you to understand about the dataset?
    \item What would have been helpful to provide additionally? Was anything missing?
    \item Do you have comments on the style of the text description? For example, was it too long, too short, too verbose?
    \item Do you have any other feedback or comments that we did not touch on today?
    \item In your opinion, what differentiates a great text description from other descriptions?
    \item Do you have any experiences with accessible data visualizations? If so, please elaborate.
\end{enumerate}

\subsubsection{UpSet Plot Questions}
\begin{enumerate}
    \item Does your mental model from the text description match the chart?
    \item Do you have any other feedback or comments now that you have seen the corresponding visualization?
\end{enumerate}

\subsection{Interview Stimuli}

\textbf{Page 1}

Introduction

For this study, you will see a text description generated for an UpSet plot. An UpSet plot is a set visualization technique similar to Venn diagrams, but unlike Venn diagrams, UpSet works for more than three sets.

Our research aims to make data visualizations more accessible to people with visual impairments. We want to understand whether text can convey similar amounts of information as a chart. We first will introduce what an UpSet plot is, and how to interpret data from the plot.
UpSet Explained

UpSet plots the intersections of sets in a table. Each column corresponds to a set. Bars at the top of the columns show the size of the sets. The row corresponds to an intersection: marks in the cells show which set is included in the intersection. The number of sets that participate in the intersection is referred to as degree. If there is no mark in any of the cells, then it is the intersection of no set, which is also referred to as the empty intersection, with a degree of 0. If there is a mark in every cell of a row, then it is the intersection of all sets.

UpSet plots the size of the intersections as bar charts to the right of the table. The table is also useful because it can be sorted in various ways. A common way is to sort by size, but it's also possible to sort by degree or sets.

Imagine an UpSet plot that shows movie data. Movies have genres like Drama, Comedy, Thriller, Mystery, or Crime. A movie can have a single genre, or it can have multiple genres. In this example, the genres are the sets. Some sets are bigger: there are more Drama movies than Mystery movies, for example. And some intersections will be more common: Thriller and Mystery might be a popular combination, while the combination of Drama, Comedy, and Thriller might be rare.
Glossary of Terms

Here are some terms we use in the text description:

Movies that don’t fall into any genre are an intersection of no set/the empty intersection.
A row with only one mark for drama (movies that are just dramas and have no other genre) is an independent set intersection.
A row with 2-3 marks like "Drama-Comedy" corresponds to a low-degree set intersection.
A row with 3-5 marks (e.g., Mystery-Crime-Thriller) corresponds to a medium-degree set intersection. A row with even more marks is a high-order set intersection.
The last case is that set containing all sets (i.e. all movie genres), which we call an all-set intersection.

That’s it for the introduction! Please stop now and ask the interviewers if you might have any clarifying questions.

Next, we’ll explore text descriptions generated for a variety of UpSet Plots. Click on the link below to see: UpSet Plot Description.

\small
\begin{lstlisting}
# Visualizing co-occurrence of CoVID 19 Symptoms with UpSet.

This is an UpSet plot that shows covid cases and their symptoms. A covid case can have multiple symptoms. The sets are covid symptoms. The items are covid cases. The intersections show how many covid cases have exactly the same symptoms. The plot shows intersections of 6 sets. All major intersections involve the set Fatigue, and Cough. The largest intersection is Anosmia, and Fatigue, with 281 elements. Other large intersections also involve Cough, Anosmia, and Fatigue. The intersection of all sets is present with 23 elements.

# Dataset Properties

The dataset contains 6 sets and 4340 elements, of which 6 sets are shown in the plot.

# Set Properties

The set sizes are diverging a lot, ranging from 148 to 1531. The largest set is Fatigue with 1531 elements, followed by Anosmia with 1051, Cough with 897, Fever with 363, Diarrhea with 350, and Shortness of Breath with 148.

# Intersection Properties

The plot is sorted by size in descending order. There are 32 non-empty intersections, all of which are shown in the plot. The largest 5 intersections are Anosmia, and Fatigue (281), Cough, Anosmia, and Fatigue (259), Just Fatigue (198), Cough, and Fatigue (179), and Just Anosmia (140).
Statistical Information

The average intersection size is 55, and the median is 24. The 90th percentile is 179, and the 10th percentile is 7. The largest set, Fatigue, is present in 78.1

# Trend Analysis

The intersection sizes peak at a value of 281 and then drastically flatten down to 1. An all set intersection is present with a size of 23. The individual set intersections are large in size. The low degree set intersections lie in the largest sized intersections. The medium degree set intersections can be seen among small and medium and large sized intersections. Among the medium sized intersections, the high order set intersections are significantly present.
\end{lstlisting}

\normalsize
\section{Text Description Examples}
\label{sec:text-descriptions}

Here we provide examples of the text descriptions used in our studies for the Movies dataset as an example of general trend.

\subsection{Movies}

Short Description no Configuration

\small
\begin{lstlisting}
This is an UpSet plot which shows the intersections of 6 sets. All major intersections involve the set Action, and Adventure. The largest intersection is Thriller, and Action, with 104 elements. Other large intersections also involve Action and Thriller. 
\end{lstlisting}
\normalsize

\noindent{}Long Description with no Configuration

\small
\begin{lstlisting}
# UpSet Introduction
This is an UpSet plot that visualizes set intersection. To learn about UpSet plots, visit REDACTED.

# Dataset Properties
The dataset contains 17 sets and 6303 elements, of which 6 sets are shown in the plot.

# Set Properties
The set sizes are diverging a lot, ranging from 68 to 503. The largest set is Action with 503 elements, followed by Thriller with 492, Adventure with 283, Children with 251, War with 143, and Western with 68.

# Intersection Properties
The plot is sorted by size in descending order. There are 28 non-empty intersections, all of which are shown in the plot. The largest 5 intersections are Just the empty inter (2569), Just Thriller (349), Just Action (218), Just Children (160), and Thriller, and Action (104).

# Statistical Information
The average intersection size is 138, and the median is 7. The 90th percentile is 218, and the 10th percentile is 1. The largest set, Action, is present in 50.0\% of all non-empty intersections. The smallest set, Western, is present in 25.0\% of all non-empty intersections.

# Trend Analysis
The intersection sizes peak at a value of 2569 and then drastically flatten down to 1. Just the empty inter is the largest by a factor of 7. The empty intersection is present with a size of 2569. An all set intersection is not present. The individual set intersections are large in size. The low degree set intersections lie in small and medium sized intersections. The medium degree set intersections can be seen among medium sized intersections. No high order intersections are present.

\end{lstlisting}
\normalsize

Long Description with Configuration

\small
\begin{lstlisting}
# UpSet Introduction
This is an UpSet plot that visualizes set intersection. To learn about UpSet plots, visit REDACTED.

# Dataset Properties
The dataset shows attributes of movie genres and ratings. The dataset contains 17 sets and 6303 elements, of which 6 sets are shown in the plot.

# Set Properties
The set sizes are diverging a lot, ranging from 68 to 503. The largest set is Action with 503 movies, followed by Thriller with 492, Adventure with 283, Children with 251, War with 143, and Western with 68.

# Intersection Properties
The plot is sorted by size in descending order. There are 28 non-empty intersections, all of which are shown in the plot. The largest 5 intersections are Just the empty inter (2569), Just Thriller (349), Just Action (218), Just Children (160), and Thriller, and Action (104).

# Statistical Information
The average intersection size is 138, and the median is 7. The 90th percentile is 218, and the 10th percentile is 1. The largest set, Action, is present in 50.0

# Trend Analysis
The intersection sizes peak at a value of 2569 and then drastically flatten down to 1. Just the empty inter is the largest by a factor of 7. The empty intersection is present with a size of 2569. An all set intersection is not present. The individual set intersections are large in size. The low degree set intersections lie in small and medium sized intersections. The medium degree set intersections can be seen among medium sized intersections. No high order intersections are present.

\end{lstlisting}
\normalsize

\section{Additional Figures}
\label{sec:addtional-figures}

\begin{figure}[ht]
    \centering
    \includegraphics[width=\linewidth]{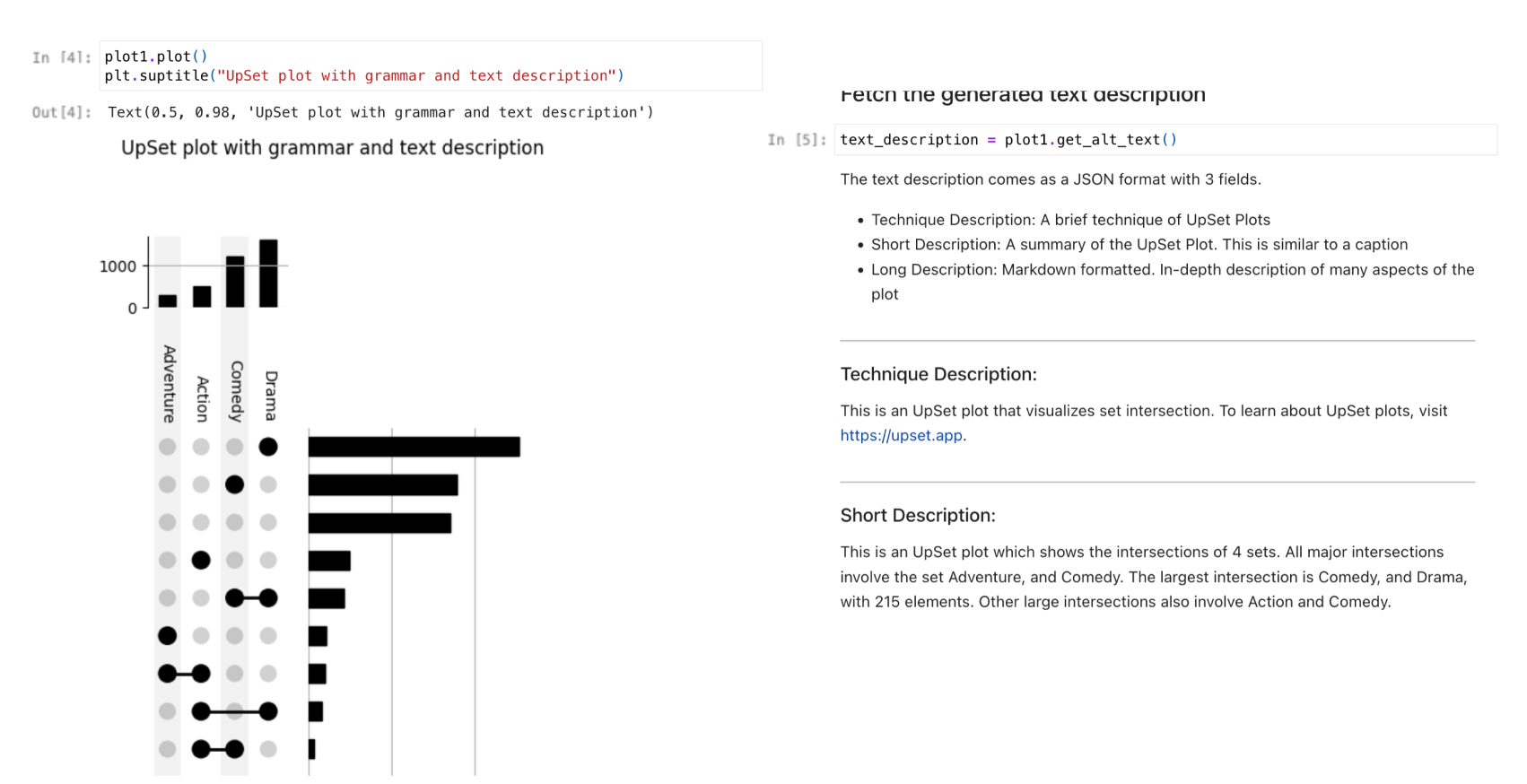}
\caption{A screen shot of a notebook using our automated text generation system. }
    \label{fig:python-usage}
\end{figure}

\begin{figure}[ht]
    \centering
\includegraphics[width=\linewidth]{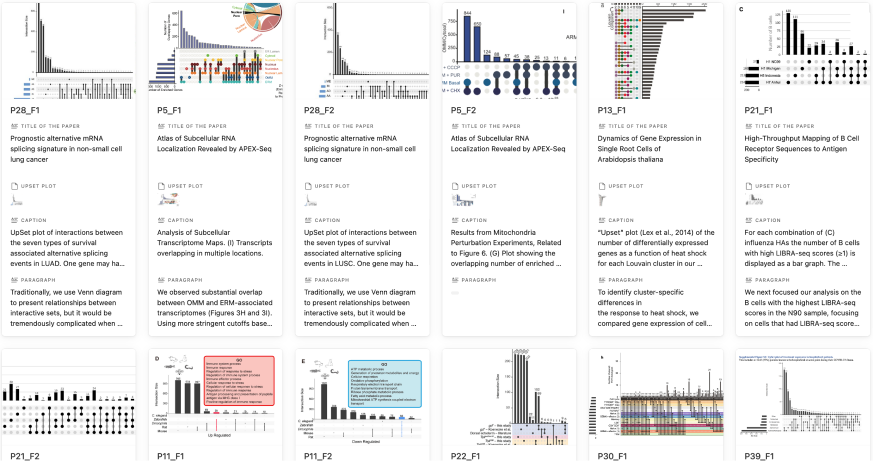}
    \caption{Selected thumbnails of UpSet plots collected for our classification of patterns found in UpSet plots.
    }
    \label{fig:upset_classification}
\end{figure}

\begin{figure}[ht]
    \centering
    \includegraphics[width=\linewidth]{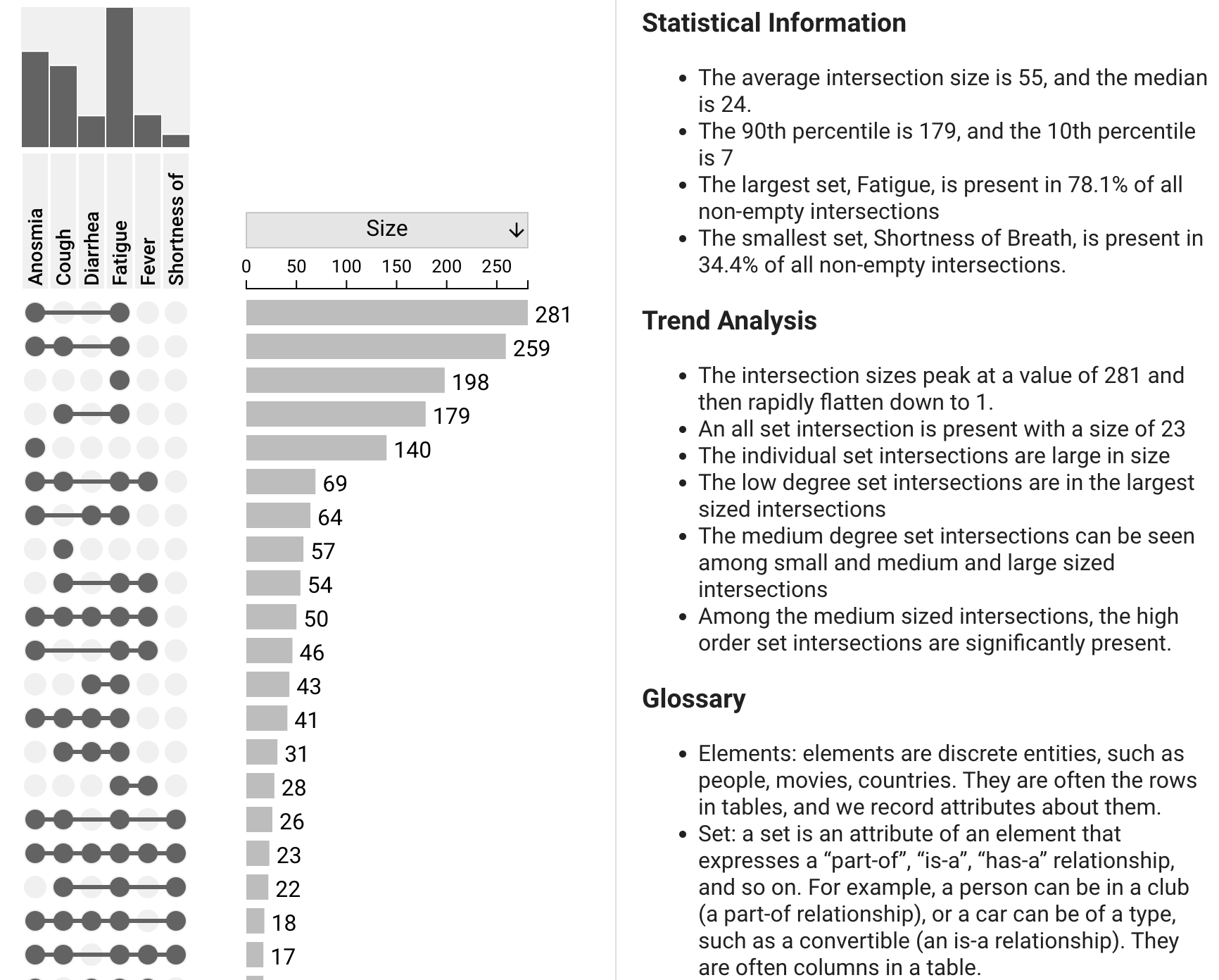}
\caption{The final version of the text description, refined based on feedback received during the interviews. While the text itself is unchanged, individual statements are listed as bullet points, and a glossary is added to the end of the description. }
    \label{fig:bullets-glossary}
\end{figure}

Finally, we include figures which that relevant to this work but were out of place in the main text.
\autoref{fig:python-usage} shows an example of our descriptive text integrated with an UpSet plot in a notebook.
\autoref{fig:upset_classification} is a snapshot of our coding of in-the-wild UpSet examples.
\autoref{fig:bullets-glossary} shows the final version of the text description, refined based on the interview results.

\end{document}